\documentclass[letterpaper,twocolumn,10pt]{article}
\usepackage[T1]{fontenc}
\usepackage{usenix2019_v3}
\usepackage[%
    breaklinks=true,colorlinks=true,linkcolor=black,%
    citecolor=black,urlcolor=black,bookmarks=true,bookmarksopen=false,%
    pdfauthor={},%
    pdftitle={},% // TODO: add title
    pdftex]{hyperref}

\usepackage[english]{babel}
\usepackage{graphicx}
\usepackage{booktabs}
\usepackage{xspace}
\usepackage[utf8]{inputenc}
\usepackage{wasysym}
\usepackage[ruled,vlined]{algorithm2e}
\usepackage{pdflscape}
\usepackage{enumitem}

\usepackage{subcaption}
\usepackage{balance}
\usepackage{float}

\usepackage[protrusion=true,expansion=true,kerning,spacing]{microtype}

\SetExtraKerning{encoding=*,font=*}{\textemdash={120,120}} % Add space around emdashes

\newcommand{\todo}[1]{}
\renewcommand{\todo}[1]{{\color{red} TODO: {#1}}}
\newcommand{\TODO}[1]{}
\renewcommand{\TODO}[1]{{\color{red} TODO: {#1}}}

\def \paragraph [#1] {\vspace{3pt}\noindent{\textbf{#1}\quad}}%

\newcommand{\proto}[1]{\textls[-30]{\textsc{#1}}\xspace}

\newcommand{\syn}{\textls[-30]{\textsc{syn}}\xspace}
\newcommand{\sa}{\textls[-30]{\textsc{syn-ack}}\xspace}
\newcommand{\rst}{\textls[-30]{\textsc{rst}}\xspace}
\newcommand{\ack}{\textls[-30]{\textsc{ack}}\xspace}
\newcommand{\fin}{\textls[-30]{\textsc{fin}}\xspace}
\newcommand{\psh}{\textls[-30]{\textsc{push}}\xspace}
\newcommand{\synack}{\sa}

\newcommand{\hostsa}{TCP-responsive\xspace}

\newcommand{\lmap}{LZR\xspace}

\title{\lmap: Identifying Unexpected Internet Services}

\begin{document}

%for single author (just remove % characters)
\author{
{\rm Liz Izhikevich}\\
Stanford University
\and
{\rm Renata Teixeira}\\
Inria, Paris\thanks{Work done while visiting Stanford University.}
\and
{\rm Zakir Durumeric}\\
Stanford University
} % end author

\maketitle

\begin{abstract}

Internet-wide scanning is a commonly used research technique that has helped
	uncover real-world attacks, find cryptographic weaknesses, and understand
	both operator and miscreant behavior. Studies that employ scanning have
	largely assumed that services are hosted on their IANA-assigned ports,
	overlooking the study of services on unusual ports. In this work, we investigate where Internet services are deployed in practice and evaluate
	the security posture of services on unexpected ports.
	%We develop a heuristic to accurately detect the presence of a real service
	%and show that middleboxes are responsible for the bulk of responses on many
	%ports.
	We show protocol deployment is more diffuse than previously believed and
	that protocols run on many additional ports beyond their primary
	IANA-assigned port. For example, only 3\% of HTTP and 6\% of TLS services
	run on ports~80 and 443, respectively. Services on non-standard ports are
	more likely to be insecure, which results in studies dramatically
	underestimating the security posture of Internet hosts. Building on our
	observations, we introduce \lmap (``Laser''), a system that identifies 99\%
	of identifiable unexpected services in five handshakes and dramatically
	reduces the time needed to perform application-layer scans on ports with few
	responsive expected services (e.g., 5500\% speedup on 27017/MongoDB). We
	conclude with recommendations for future studies.

\end{abstract}

\section{Introduction}

Internet-wide scanning---the process of connecting to every public IPv4 address
on a targeted port---is a standard research technique for understanding
real-world service configuration and deployment.  Leveraging tools like
ZMap~\cite{durumeric2013zmap} and Masscan~\cite{graham2014masscan}, more than
300~papers have used Internet-wide scanning to discover weaknesses in TLS, SSH,
and the Web PKI~\cite{amann2017mission, holz2015tls, checkoway2014practical,
aviram2016drown, beurdouche2015messy, durumeric2013analysis,
holz2011ssl,brubaker2014using, adrian2015imperfect, heninger2012mining}, to
uncover real-world attacks~\cite{marczak2014governments,
rossow2014amplification, durumeric2015neither}, and to better understand
botnets~\cite{mirai, hunting}, ICS/IoT deployment~\cite{costin2014large,
mirian2016internet, springall2016ftp}, censorship~\cite{pearce2017augur,
pearce2017global, khattak2016you}, and operator behavior~\cite{li2016you,
durumeric2014matter, durumeric2014internet}.

Past scanning studies have largely assumed that services are hosted on their
IANA-assigned ports (e.g., HTTPS on TCP/443) and have overlooked scanning
additional ports for unexpected services. Yet, many of these same studies have
also observed that a non-negligible fraction of the hosts that respond to a \syn
scan never complete the expected application-layer
handshake~\cite{mirian2016internet, springall2016ftp, heninger2012mining,
durumeric2013zmap, durumeric2013analysis, durumeric2015search}. It is unclear
whether operators hide services on unexpected ports, whether scanners fail to
account for protocol inconsistencies or server-side implementation errors, or
whether firewalls detect scanning and block further interaction. In this work,
we investigate where Internet services are deployed in practice, and we evaluate
the security posture of services hosted in unexpected places.
%Moreover, it is unclear whether \syn scanning is even accurate enough to
%approximate service liveness, as, on many ports, these discrepancies affect the
%majority of hosts.  If the security community is to continue using Internet
%scanning to make technical and policy decisions, it behooves us to accurately
%understand the IPv4 service ecosystem and to find a more accurate and efficient
%way to find and study unexpected services.

%This discrepancy varies dramatically by port; on TCP/80,
%87\% of hosts that responded to a \syn scan will actually respond to an HTTP
%\texttt{GET\,/} request but only 4\% of TCP/102 hosts speak Siemens S7.
%This is not surprising---scanners like ZMap assume that targeted services
%(e.g., SSH) are deployed on a single port.
%live on the targeted port (e.g.,
%TCP/22). Otherwise, to achieve full coverage of listening services, one must
%scan all 65,535~ports across 3.7~billion IPv4 addresses and attempt an unbounded
%number of application-layer handshakes on each port. At 10\,gbE line rates,
%doing so would take at least 7.5~months~\cite{adrian2014zippier}.  Consequently,
%the unexpected services ecosystem and its security posture is completely
%unknown.
%
%, and an estimated 4.5M hosts on TCP/21 (FTP) stop responding after sending a
%\synack.\looseness=-1 and on TCP/443, 80\% complete a TLS handshake.  On less
%popular protocols, only a fraction of hosts complete a handshake:  43\% of
%TCP/23 hosts speak Telnet and 4\% of TCP/102 hosts speak Siemens S7. On TCP/21
%(FTP), an estimated 4.5M hosts stop responding after sending a
%\synack.\looseness=-1

We start by investigating services that do not appear to speak the expected
IANA-assigned protocol. We confirm that up to 96\% of services (by port) do not
complete the expected application-layer (L7) handshake on 37~popular ports
(Section~\ref{sec:l4}).  We introduce a heuristic that infers server-side TCP
state, which we use to show that 28\% of initially-responsive services do not
allow any L7 data exchange. Rather, 12\% immediately tear down the connection,
5\% prevent an L7 handshake by specifying a zero TCP window, 0.6\% are blocked
from receiving our \ack, and 11\% ``shun'' our IP between the discovery and
application-layer scan phases. We trace these behaviors to middleboxes and
firewalls, and we evaluate their efficacy at enabling scan evasion.
%Network appliances implement \syn cookies~\cite{syncookieslemon2002resisting}
%on every port to prevent DDoS attacks; they \synack for all ports and check for
%a backend service only after receiving an \ack; the vast majority host no
%service. We show that scanners like ZMap should verify that servers acknowledge
%data before opening an OS TCP socket rather than rely on a single \synack; 45\%
%of services that \sa data complete the expected L7 handshake.

While network defenses account for most L7 unresponsive services, a significant
number of services are TCP compliant, but fail the expected L7 handshake (e.g.,
14\% on TCP/80 and 96\% on TCP/102). We show that this is due to services
running on unexpected ports, protocol handshakes that require pre-established
secrets, and network-based protections that acknowledge data on every port but
speak no detectable protocol (Sections~\ref{sec:l7}--\ref{sec:optimizations}).
Notably, protocol deployment is exceptionally diffuse. For example, only 3.0\%
of HTTP and 6.4\% of TLS services run on ports~80 and 443, respectively.
Achieving 90\% coverage of TLS-based services requires scanning 40K~ports.
Worryingly, services deployed on unexpected ports have worse security postures,
which we trace back to IoT devices that host insecure services on non-standard
ports.
%because many of the hosts with services on unexpected ports are IoT devices.
%
%We further discover that unexpected services are often even greater
%contributors to the Internet's vulnerability than expected services, due to
%poor security practices (e.g., 1.17~times more TLS certificates have a known
%private key on unexpected ports---over twice as much than reported in previous
%work~\cite{heninger2012mining,hastings2016weak}) and the prevalence of IoT
%devices (e.g., 50\% of TLS on unexpected ports belong to an IoT device), which
%frequently exhibit weak security practices.
%
%While services are often found on unexpected ports, the improvement from trying
%additional handshakes diminishes quickly; 95\% of services are found in the
%first 10~handshakes.

To enable researchers to more comprehensively find Internet services, we
introduce \lmap (``Laser''), a system that efficiently filters hosts that do not
speak any L7 protocol and identifies unexpected services
(Section~\ref{sub:l4_heuristic}). \lmap can fingerprint 88\% of identifiable
services with a single packet and 99\% of identifiable unexpected services with
five handshakes. \lmap also speeds up scans by quickly filtering the bulk of
seemingly-responsive hosts that \synack but cannot complete an application layer
handshake. For example, on port~27017, \lmap filters out 80\% of hosts that
\synack, decreasing the time to complete scans of MongoDB by 55~times, while
still identifying 99.6\% of MongoDB services and identifying an additional
23K~hosts running unexpected protocols (a 31\% coverage increase for the port).

%On less popular ports, \lmap shortens application-layer scans
%
%by upwards of 40~hours by removing
%
Our work concludes with recommendations for future studies. We hope that by
shedding light on the ecosystem of unexpected services, and by releasing \lmap
as an open-source tool, we enable security researchers to more accurately
understand Internet services.

\section{Identifying Real TCP Services}
\label{sec:l4}

Fast research scans of the Internet are typically conducted in two phases
today~\cite{durumeric2013zmap, heninger2012mining, holz2011ssl,
durumeric2015search}. In the first stage, a scanner like
ZMap~\cite{durumeric2013zmap} statelessly sends \syn packets to public IPv4
addresses. Then, in a second process, a stateful scanner like
ZGrab~\cite{durumeric2015search} performs complex follow-up handshakes using the
kernel TCP/IP stack. The two-phased nature of Internet scanning is largely
attributable to ZMap's architecture, which uses a stateless network stack to
efficiently probe services, but is unable to complete handshakes that require
maintaining local state. The biases and unintended consequences from scanning in
two phases have not been investigated, and worryingly, prior studies have
repeatedly noted that more than half of the IPv4 hosts that respond to a SYN
scan never complete a follow-up application-layer handshake
(e.g.,~\cite{mirian2016internet, springall2016ftp, heninger2012mining,
durumeric2013zmap, durumeric2013analysis}).

In this section, we investigate this discrepancy. We show that TCP liveness does
not accurately indicate the presence of an application-layer service due to
several common security protections, including middleboxes and user-space
firewalls. Guided by TCP's design~\cite{tcp}, we uncover five defensive
behaviors that degrade the signal provided by L4 responsiveness.  We quantify
the deployment of these defenses, and we evaluate their efficacy at protecting
against DDoS attacks and evading Internet scans. We then go on to develop a
better L4 heuristic to approximate application-layer liveness, which we use to
better understand service deployment in Section~\ref{sec:l7}.

\subsection{Layer 4 versus Layer 7 Liveness}
\label{sub:sec:twophase}

We start our investigation by confirming whether TCP-responsive hosts (i.e.,
hosts that reply with a \sa packet) complete the IANA assigned~\cite{ianaports}
application-layer handshake. Mimicking prior Internet scans
(e.g.,~\cite{heninger2012mining, amann2017mission, adrian2015imperfect,
zhang2014mismanagement, checkoway2016systematic}), we perform a two-phase scan
in which we send a \syn packet to a random 1\% sample of public IPv4 addresses
using ZMap~\cite{durumeric2013zmap} and immediately attempt a follow-up
application handshake using ZGrab~\cite{durumeric2015search}. We scan all
IANA-assigned ports with available ZGrab scanners (i.e., 37~ports in
Appendix~\ref{app:protos}) on November 12--14, 2019. We follow the best
practices set forth by Durumeric et~al.~\cite{durumeric2013zmap} to minimize
scan impact, and we exclude networks that have previously contacted us. We
receive no complaints, but note that we have used our network in the past for
other experiments and exclude operators who previously requested
removal.\looseness=-1

\begin{figure}[h]
  \centerline{\includegraphics[width=\linewidth]{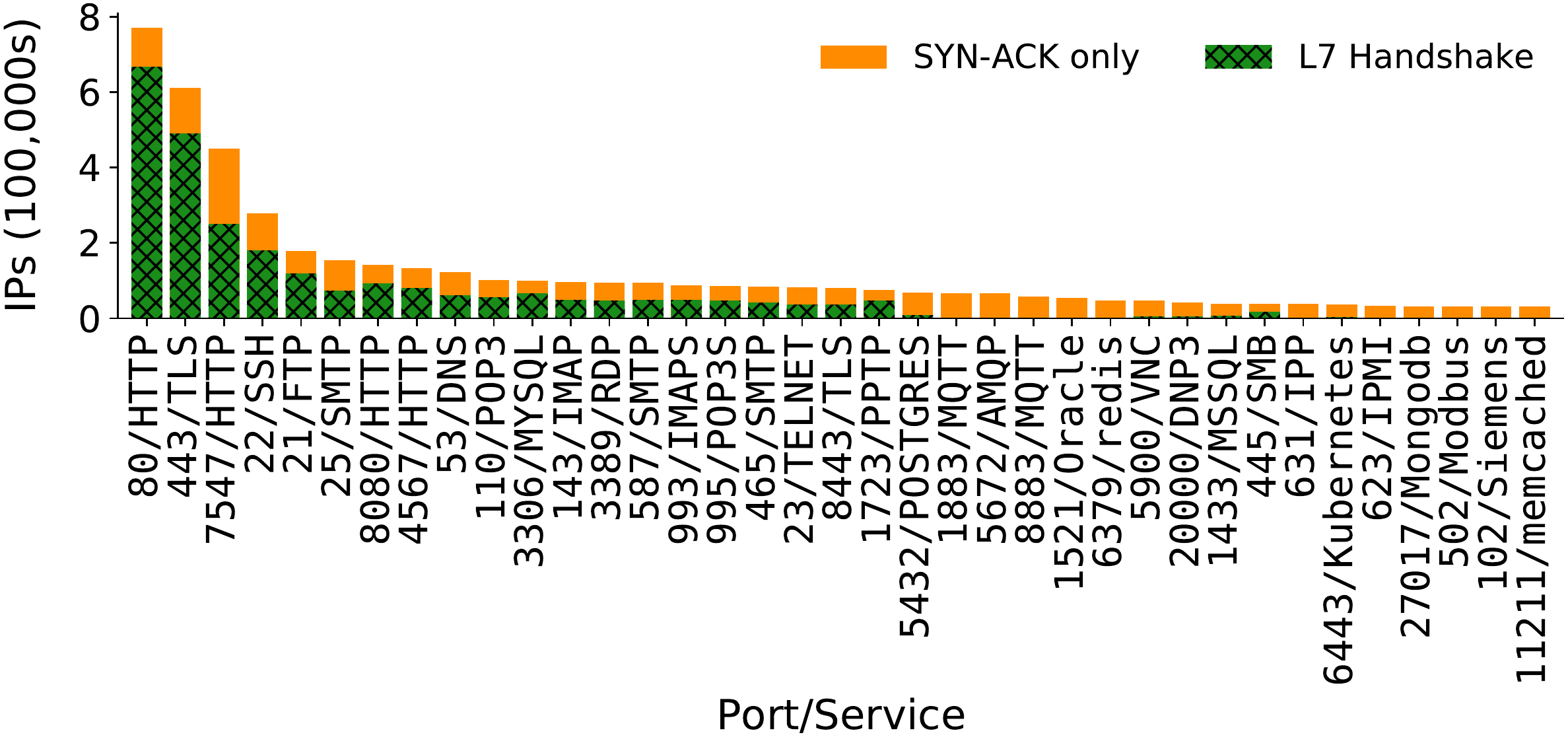}}
  \caption{\textbf{L4 vs.\ L7 Responsiveness}---%
	A significant fraction of hosts that respond with a \synack
	packet never complete the expected application-layer handshake. The
	difference varies dramatically across ports by both percent difference (14--96\%) and
	raw count (21,050--200,902).
  }
  \label{fig:sa_v_zgrab}
  \vspace{-10pt}
\end{figure}

Consistent with prior studies~\cite{mirian2016internet, springall2016ftp,
heninger2012mining, durumeric2013zmap, durumeric2013analysis}, we find that a
considerable fraction of \hostsa hosts never complete the expected L7 handshake
(Figure~\ref{fig:sa_v_zgrab}).
The raw number of L7-unresponsive hosts varies from
21K~unresponsive hosts on 502/Modbus to 201K~hosts on 443/HTTPS ($\mu=54,542$,
$\sigma^2=31,002$). We see this heavy-tail distribution throughout our
investigation and we present our results for both popular and unpopular ports.
We split ports into the two categories using Grubbs's test for
outliers~\cite{grubbs1950sample} with a 99.9\% confidence interval based on the
total number \sa{s} and the presence of an expected service. Our popular set
contains ports 80, 443, 7547, 22, 21, and 25; the unpopular set contains the
remaining 31~ports.
Popular protocols are most likely to complete the expected L7 handshake:\footnote{Spearman's Correlation p-value of port rank (based on number
of \sa) relative to L7 and \sa percent difference is $5\times 10^{-11}$.} 86\%
and 80\% of \hostsa hosts on ports~80 and~443 complete an HTTP(S) handshake
while only 9\% and 4\% of hosts on ports~502 and 102 speak Modbus and Siemens S7
(two SCADA protocols).

In the following section we start our investigation of L7-unresponsivess by analyzing the changing state of services between the two phases of scanning.

\subsection{Connection Shunning}
\label{sub:sec:db1}

About 1.6\% of services on popular ports and 5\% of services on unpopular ports do
not respond with a \sa during our follow-up ZGrab TCP handshake. This could be
due to DHCP churn, transient network failure, or the destination host blocking
the scanner between handshakes (``connection shunning''). To determine whether
hosts ``shun'' scanners, we connect to \hostsa hosts found by ZMap from two IP
addresses: the original IP address used by ZMap to identify the host and a fresh
IP that has not previously contacted the host. We scan a random ephemeral port,
48302, because we see the largest fraction of disappearing hosts on unpopular
ports. We find that 70\% of IPs that do not respond a second time on the used IP
do respond to the fresh IP, indicating that most hosts that go missing between
scan stages are typically not lost due to churn or network failure.

In the case that the fresh IP receives a \sa, we observe two types of responses
from the previously-used IP: no response (93\%) and \textsc{rst} packet (7\%).
This blocking occurs at the IP granularity: once a scanner has been blocked by a
host, the host will not respond with a \sa on any port.  We further confirm that
connection shunning is not a defensive reaction---triggered by failing to complete
an application layer handshake---by running a 1\% IPv4 scan of all popular ports
using ZGrab for the initial host discovery. The same fraction of connections are
shunned as when ZMap is used.

\begin{figure}[h]
%	\begin{minipage}[t]{0.49\textwidth}
  \includegraphics[width=\linewidth]{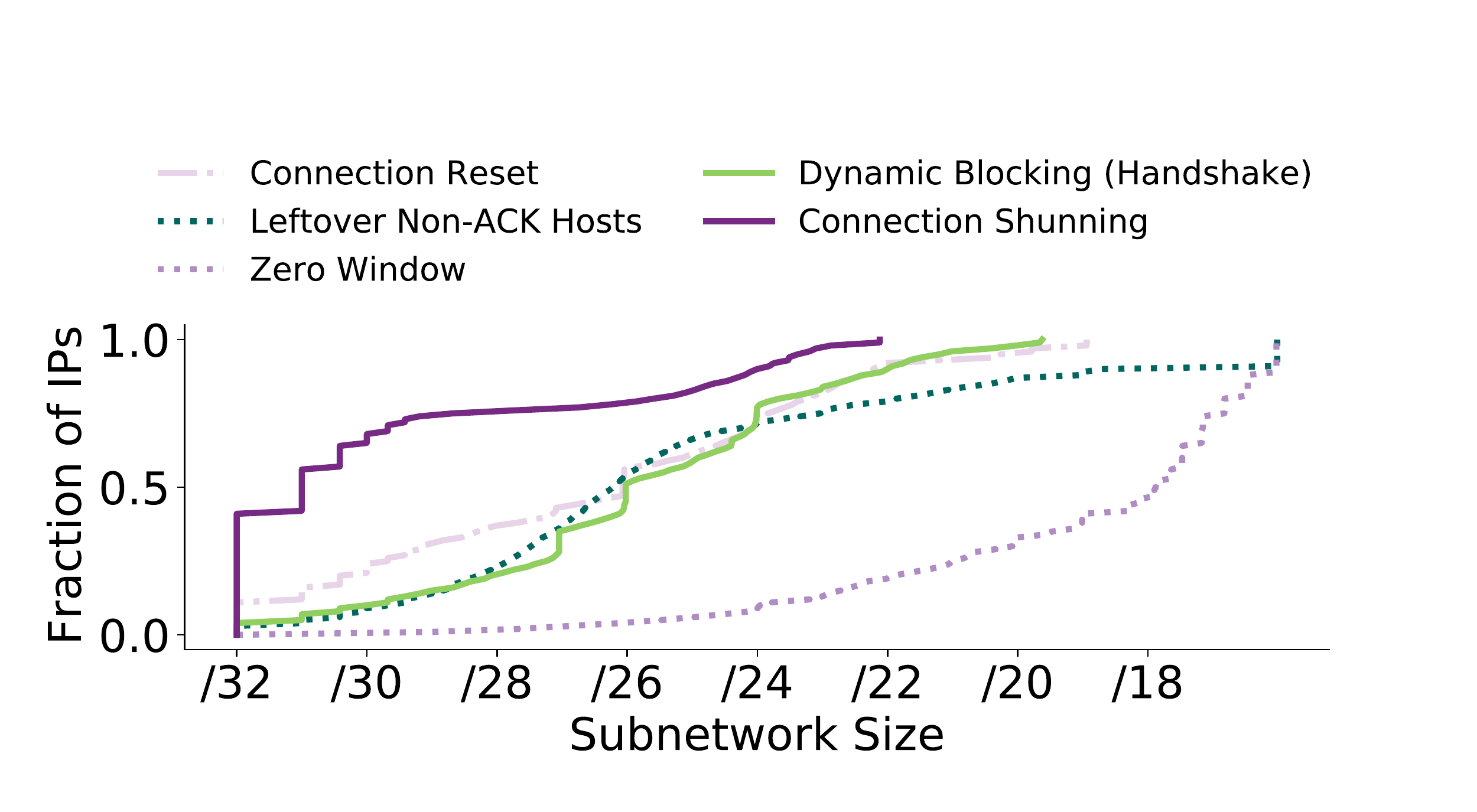}
  \caption{\textbf{Network Granularity of TCP Blocking}---%
  Some protections appear to be host-based while others are more prevalent on
	large networks. Zero Window DDoS protections are most likely to appear at
	a large network granularity, while connection shunning is more likely a
	host-level behavior.\looseness=-1
  }
  \label{fig:win0_cdf}
%  \end{minipage}
\end{figure}

We find that connection shunning is deployed at both the host and network
granularity by computing the largest blocks of consecutive TCP-Responsive IPs
that show shunning behavior on a random ephemeral port: 40\% of networks that
shun scanners are /32s (i.e., individual hosts) and 10\% of IPs block in groups
larger than a /24 (Figure~\ref{fig:win0_cdf}). The largest network to deploy
connection shunning is a /20 owned by Alestra Net (ASN\,11172), a Mexican ISP\@.

Both network hardware (e.g., Cisco IOS-based
routers~\cite{headquarters2012security}) and host software (e.g.,
Snort~\cite{roesch2002snort}) document connection shunning and dynamic blocking
as features where connections are blocked after an IP is classified as
malicious. Connection shunning prevents clients from using a single source-IP to
scan the network and forces scanners to use multiple source IPs to reach the
end-host, thereby dramatically increasing the cost for an attacker. We compare
the number of legitimate services found when using both single and multiple
source-IPs during scanning and find no evidence that any hosts that shun
connections host legitimate services. We thereby conclude that they can be
safely ignored in security studies if they can be efficiently filtered.

\subsection{Do TCP-Responsive Hosts Speak TCP?}
\label{sub:sec:tcp_state}

\begin{figure}[h]
  \centering
  \includegraphics[width=\linewidth]{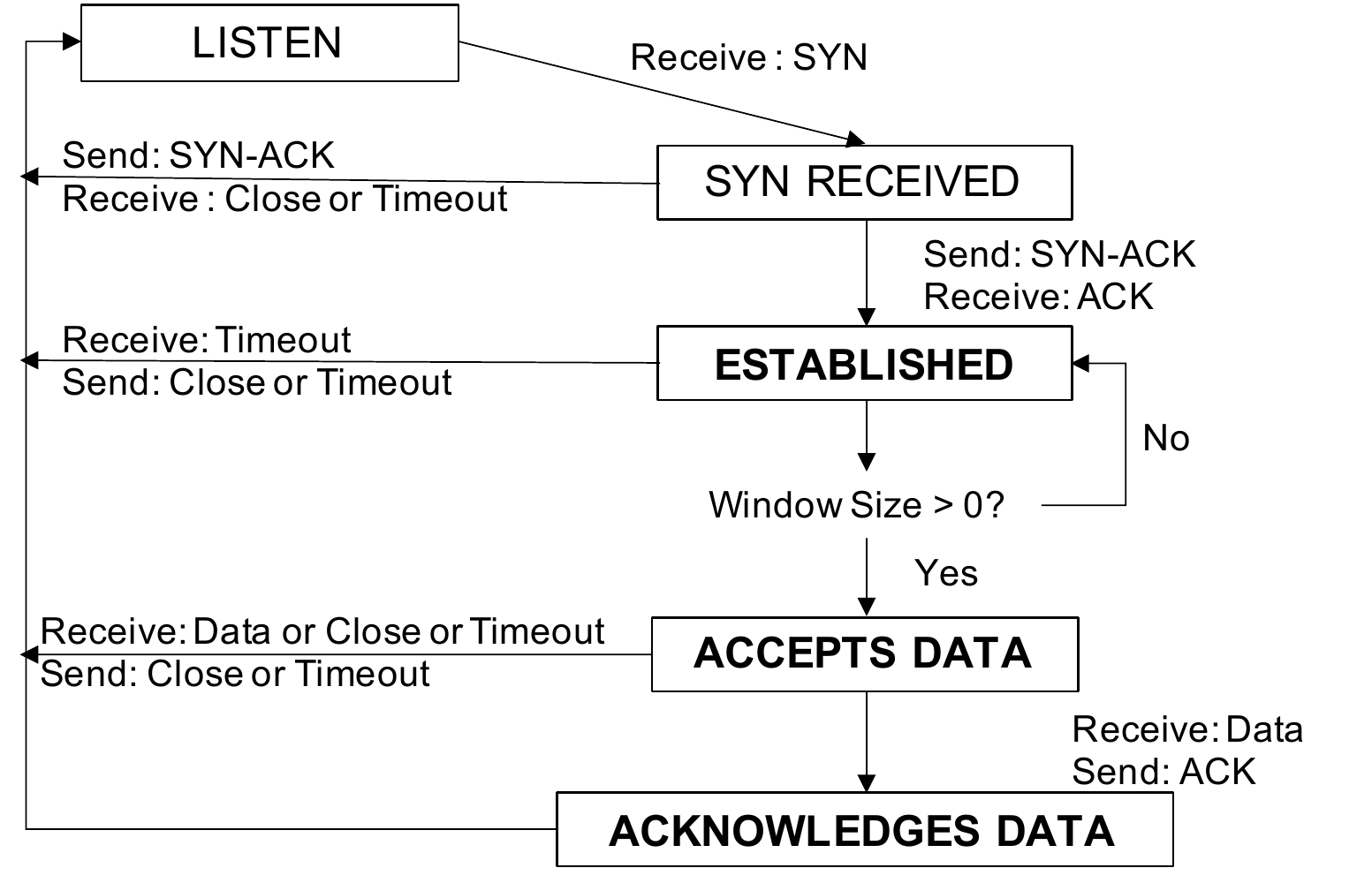}
  \caption{\textbf{Client Perspective of Server TCP State}---%
	We investigate L7 service liveness based on a modified version of the TCP
	state machine in RFC\,793~\cite{tcp}. We introduce two new states: ``accepts data'' and ``acknowledges data'' because
	an established connection cannot necessarily exchange data.
  }
  \label{fig:flowchart}
\end{figure}

The vast majority of services (average of 96\% across ports) that do not
complete an application-layer handshake respond with a \sa during the second
(ZGrab) handshake. In the remainder of the section, we explore whether these
hosts reach a state where they can exchange application-layer data or simply
stop responding after sending a \sa. In Figure~\ref{fig:flowchart}, we provide a
modified TCP state diagram based on RFC\,793~\cite{tcp} that captures what a
scanner can infer about a server's TCP state, which we use to guide our
investigation. For a TCP connection to enter the \proto{established} state, the
server sends only a single packet (\sa). Once the client has sent an \ack, it
can normally send data---the amount specified by the server window size in the
\sa packet.

We note that TCP has an edge case in which the server can respond with a
zero-sized window in its \sa~\cite{tcp}. In this situation, the client is
expected to send follow-up \ack packets to probe when the server is ready to
accept data. We add a new \proto{accepts data} state in
Figure~\ref{fig:flowchart} to capture whether a server is ready for data. Once
the server has reached the \proto{accepts data} state, it is expected to keep
the TCP connection open long enough to receive data and to acknowledge receipt.
We define \proto{acknowledges data} as the server allowing the client to send
data and acknowledging client data.

\begin{algorithm}[h]
\small
\SetAlgoLined
%\KwResult{Checks if host acknowledges data}

Send \syn \\
\If{ receive \rst or \fin or Timeout } {
    return NO\_ACK\_HOST
}

\tcp{checking for zero window sizes}
Print syn-ack.window\_size\\
\tcp{sending protocol-agnostic data}
Send "\textbackslash{n}\textbackslash{n}" \\

\tcp{Time for 8 re-transmissions (RFC 1122 rec.)}
\While{ timeout < 100 seconds }{

\If{ received \ack  }{
    return ACK\_HOST
}
\If{ received \rst or \fin }{
    return NO\_ACK\_HOST
}
}
{return NO\_ACK\_HOST} \tcp{host has timed out}

  \caption{Deducing Server TCP State}
  \label{fig:alg_1}
\end{algorithm}

\begin{figure*}[]
  \centering
  \begin{subfigure}{\textwidth}
    \centerline{\includegraphics[width=\linewidth]{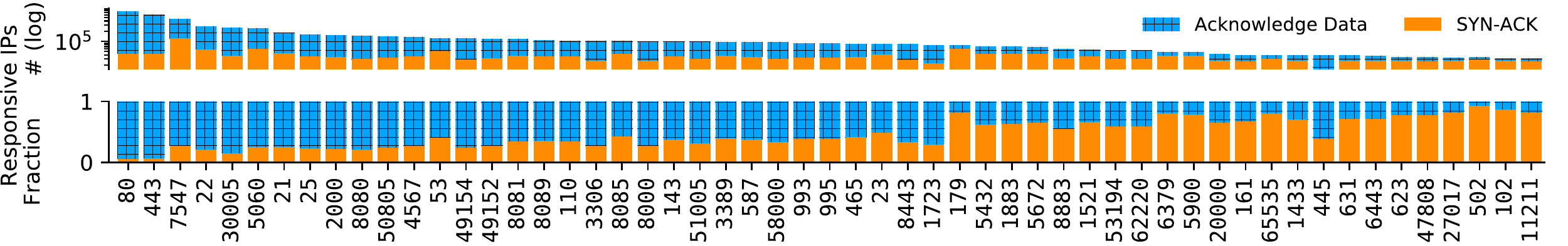}}
  \caption{Portion of \hostsa hosts that fail to acknowledge data
  }
  \label{fig:l4_all}
  \end{subfigure}
  \hfill
  \begin{subfigure}{\textwidth}
  \centerline{\includegraphics[width=\linewidth]{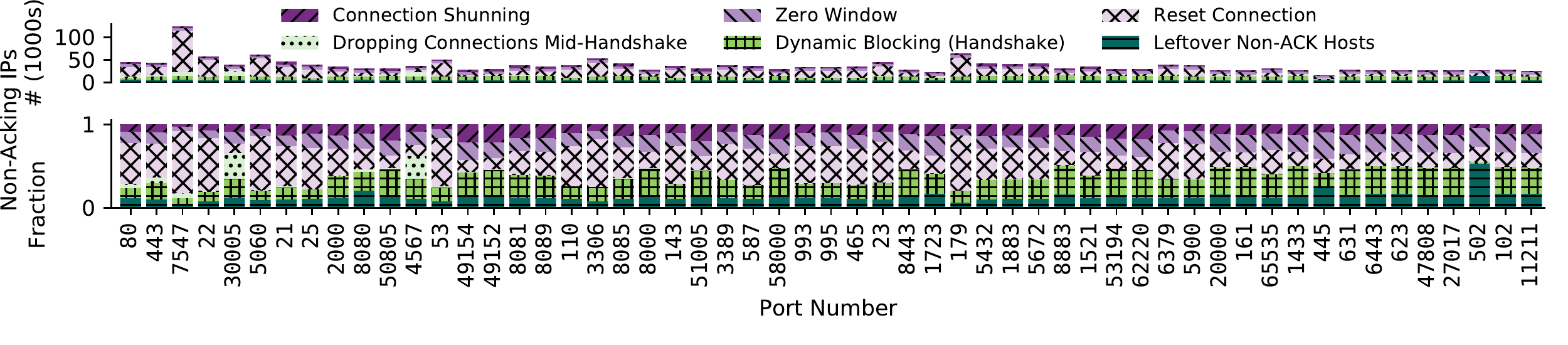}}
  \caption{Reasons \sa-only hosts fail to acknowledge data}
  \label{fig:l4_no_establish}
  \end{subfigure}
  \caption{\textbf{Unexpected TCP Behavior of IPv4 Hosts}---%
	An average 16\% of services on popular ports and 40\% of services on unpopular ports that respond to a TCP \syn scan with a
	\sa packet do not fully speak TCP\@. Here, we show the portion of hosts by
	port that do not acknowledge client data and the breakdown of reasons why.
  }
\end{figure*}

To test how far into a TCP session servers reach, we develop a new
 scanner based on ZGrab~\cite{zgrab} that establishes a TCP connection, sends two
newlines, and deduces the server TCP state (Algorithm~\ref{fig:alg_1}). We scan
random 1\% samples of IPv4 addresses on a random 2,000~ports as well as the
37~IANA assigned ports that host protocols with ZGrab scanners
(Appendix~\ref{app:protos}). An average 16\% of services on popular ports and
40\% of services on unpopular ports fail to acknowledge data
(Figure~\ref{fig:l4_all}). We detail why in the remainder of this section.

% As we will
%discuss in Section~\ref{}, hosts that do not acknowledge data account for about
%11--53\% of the hosts that do not speak the expected application-layer
%protocol.

%In the remainder of this section, we investigate which L4 behaviors cause a host to fail to acknowledge
%data and what the L4 behavior is attributed to (e.g., a particular middlebox). Concretely, we use the results of our new ZGrab scanner to discover and cluster all exhibited L4 behaviors which inhibit the acknowledgement of data.
%We show a high-level breakdown of root causes in
%Figure~\ref{fig:l4_no_establish}. As an aside, we note that a connection can fail to reach
%the established state if the server sends a \rst packet after sending a \sa,
%tearing the connection down before the client can send an \ack. This happens
%less than 0.1\% of the time across all ports and we exclude this case from our
%discussion.

%% TODO: This doesn't fit in here. move to later.
%Interestingly, several
%security-relevant ports diverge from this pattern: an anomalously small fraction
%of \hostsa hosts acknowledge data on port~179 (BGP), but on port~445, an
%anomalously large fraction of hosts do.  \rt{where do you explain what is going
%on for 179 and 445? } IPs that respond on port 502 (Modbus), have an 85\%
%likelihood of responding on port~179 (BGP), \TK times as likely as a random IP
%address responding on port~179, which we discuss in Section~\ref{TK}.

\subsection{Zero Window DDoS Protections}
\label{sub:sec:syn_sent}

Of the services that never acknowledge data, 13\% of services on popular ports
and 26\% on unpopular ports actively prevent clients from sending data by
specifying a zero-sized TCP window and never increasing it. Across all scanned
ports, at least 99.94\% of hosts with a zero window never increase it; 90\% do
not respond to secondary probes and 10\% reset the connection. The behavior
appears to be network- or host-based rather than service-based: 99\% of hosts
that respond with a zero-window on one port will send a zero-sized window on all
ports. Offhand, this behavior appears self-defeating. Hosts that respond and
never increase window size might as well never respond. However, we find the
feature in a Juniper networks patent~\cite{juniper_patent} and used in Juniper's
Secure Service Gateway Proxy~\cite{juniper} to prevent DDoS attacks through
network-based SYN cookies. The protection responds to all \syn packets with a
zero-window \textsc{syn-ack}\@. Once the client completes the three-way
handshake by sending an \ack, the firewall sends a \syn packet to the backend
server to establish the connection. By maintaining a zero-sized TCP window with
the client, the middlebox prevents the client from sending data it cannot yet
forward to the backend server.

%\begin{figure*}[h]
%  \begin{minipage}[t]{0.49\textwidth}
%  \includegraphics[width=\linewidth]{figures_paper/cdf_numases_all.pdf}
%  \caption{\textbf{CDF ASes that account for TCP blocking}---Most L4 behaviors occur in a %relatively small number of ASes.
%  \todo{fix line width}
 %
%\todo{make height shorter}
%\todo{patterns}
%\todo{complementary CDF ? want to see different patterns on tail }
%\todo{split purple in this plot}
%\todo{in caption say that y -axis is fraction of IPs per category}
%\todo{change window size 0 : Zero Window}
%\todo{change to cumulative fraction of IPs if CCDF does not work out}
%\todo{make these lines less thick and add one black slightly slicker that
%		captures \emph{all} non-acknowledging hosts}
%}
%\label{fig:ases_cdf}
%\end{minipage}
%\hfill

Zero-window \sa{}s are deployed across entire sub-networks: 90\% of IPs that \sa with
	a zero window do so in a network larger than a /24
	(Figure~\ref{fig:win0_cdf}). The largest network, the State of Florida
	Department of Management Services (ASN\,8103), is responsible for 16\% of
	all zero-windows Internet-wide and accounts for around 3\% of
	\emph{all} \sa{}s on a random port. The TTL for \sa is consistently one hop
	closer than the later \rst, further confirming a network appliance is
	responsible.

\subsection{Dropping Connections Mid-Handshake}
\label{sub:sec:cont_sa}

Beyond specifying a zero window, an average 2\% of the hosts per port that never
	acknowledge data do not appear to complete a three-way handshake, despite the
	client sending an \ack (Figure~\ref{fig:l4_no_establish}). We infer that the
	server never reaches the \textsc{established} state based on a continual
	stream of \synack packets (average 7.8~\synack re-transmissions).
	Hosts do not simply have broken TCP
	stacks; in the case of MCI Communication Services, for example, IPs that
	re-transmit \sa{}s on port 4567 have compliant behavior on other
	ports (e.g., RDP on TCP/3389). Real services respond with a TTL over twice
	as large as the TTL value which re-transmits the \sa, suggesting that a
	middlebox selectively drops packets. Dropping connections mid-handshake is a
	defensive behavior exhibited primarily by ISPs protecting consumer premise
	equipment: CenturyLink (AS\,209), Frontier Communications (AS\,5650), and
	MCI Communications Services (AS\,701) all drop inbound traffic to port
	4567/TRAM post-SYN (accounting for 96\% of dropped connections). Korea
	Telecom (AS\,4766) and Axtel (AS\,6503)---accounting for 73\%---interrupt connections on 7547/CWMP. The behavior is rare on common ports (e.g., only 5\%
	of TCP-responsive hosts that do not acknowledge data drop connections
	mid-handshake on port 80).

\subsection{Reset Connections}

An average 73\% of services on popular ports and 34\% of services on unpopular
ports that do not acknowledge data reach the \proto{established} state but will
immediately reset the connection after the client completes the three-way
handshake (Figure~\ref{fig:l4_no_establish}).  Per RFC\,793~\cite{tcp}, if a
server does not want to communicate with a client (e.g., due to mismatches in
``security clearances''), the server should close the TCP connection after the
client acknowledges the \sa. This is also how user-space firewalls like
DenyHosts~\cite{denyhosts} appear to scanners. While we cannot detect what
software closes a connection, we note that networks that \rst on port 22 are
10~times more likely to do so in block-sizes of /32 than port~80, implying that
blocking happens more often on hosts running SSH compared to HTTP, consistent with Wan et~al.'s findings~\cite{wan20origins}.
Network-level behavior looks to be caused by DDoS protections similar to the
networks that send zero-window \sa{}s. To protect against
\syn-flooding, middleboxes send a \sa on behalf of the server and later
establish a connection with the server after the client has finished the
three-way handshake. If the server refuses the connection, the middlebox
terminates the client connection. This functionality is available in Cisco
IOS-based routers as a part of their threat detection logic~\cite{ciscoshun}.

%Reset connections are seen at both the host- and network-level granularity.
%Over 50\% of IPs that reset a connection on port~49227 will do so in a group larger
%than a /26 (Figure~\ref{fig:win0_cdf}), but 13\% of hosts on port~49227 close a TCP
%connection after a handshake when their neighbor does not.
%On port~80, responsive hosts reset the connection in 8\% of ASes. On a random
%ephemeral port, 49227, there is at least one IP which sends a \rst after a
%handshake in 20\% of all responding ASes (1115~ASes in total).

The behavior is visible in prominent networks, with more than 40\% of such IPs
located in Korea Telecom, Vodaphone Australia, OVH, and Akamai.  Hosts
are 20\% more likely to close a connection on popular ports because Google load
balancers in AS\,19527 come with a standard firewall policy that accept traffic
on these ports by default---in order to be able to perform service health
checks---and rely on the backend virtual machine to reset connections if the
port is closed~\cite{google-stackovergflow,google-loadbalance}.

\subsection{Dynamic Blocking after Handshake}
\label{sub:sec:no_ack}

Not all hosts that fail to acknowledge data send \textsc{rst}s or continually
re-transmit \sa{}s. Many simply never acknowledge any data. An average of 10\%
services on popular ports and 18\% of services on unpopular ports do not
acknowledge client data (Figure~\ref{fig:l4_no_establish}).  These hosts
frequently do not respond to later follow-up handshakes either. This
``shunning'' behavior is similar---but not identical---to the behavior we found
in Section~\ref{sub:sec:db1} and has previously been documented in the Great
Firewall of China~\cite{claytonChinaFirewall} where it is used to stop future
connections, triggered only when data is sent.
%In this shunning situation, it is likely not that the host fails to speak TCP
%but rather that the server stops responding after sending a \sa.

To differentiate between hosts that shun the scanner after a handshake from
those that simply never acknowledge data, we simultaneously attempt an L7
handshake with initially-responsive hosts that did not acknowledge data from two
IP addresses, one that matches the initial connection and one that differs. Of
the initially unresponsive IPs, 98\% respond to the fresh IP,
indicating the behavior is not likely due to transient network failure, but
rather explicit blocking of incoming connections. In total, post-handshake
dynamic blocking accounts for 6\% and 12\% of the remaining hosts that do
not acknowledge data for common port and uncommon port hosts respectively.  Note
that this behavior only occurs after a three-way handshake, thereby differing
from connection shunning (Section~\ref{sub:sec:db1}).  The largest network to
dynamically block after a handshake is Coming ABCDE HK (AS\,133201), which
accounts for 48\% of all IPs that block after a handshake. We also discover
a similar TTL phenomenon as described in Section~\ref{sub:sec:syn_sent} implying
a middlebox-based protection.
%Dynamic blocking after a handshake is a more popular form of dynamic
%blocking---more than twice as many IPs only block after a full
%handshake attempt and is seen in up to 5\% more unique ASes. 50\% of IPs
%dynamic blocking after a handshake do so in a group larger than /26.

We deduce that the rest of the hosts that fail to acknowledge data are not
performing dynamic blocking because though they will not respond to anything
after the actual handshake, they do consistently respond to all scans (no matter
the source IP). Vodaphone (AS\,133612) and Webclassit (AS\,34358) have this
behavior across all scanned ports and make up 66\% of all IPs with such a
behavior. We find similar evidence of mismatching TTL values, which indicate a
middlebox.

\subsection{Efficacy of Middlebox Protections}
\label{sub:sec:efficacy}

Identifiable middlebox protections are common. About 16\% of the services on
popular and 40\% of the services on unpopular ports that respond to a \syn
packet---but do not speak any identifiable L7 protocol---are artifacts of DDoS
and scanning protections; 40\% of routed ASes contain at least one such
protection. Reset connections after a handshake---a behavior found in software
like DenyHosts~\cite{denyhosts}---is by far the most common behavior by both IP
and AS, and is present in 34\% of ASes. Middleboxes employing connection
shunning or dynamic blocking are each used by 6\% of networks, and Juniper's
patented zero-window DDoS protection appears in 2\% of networks.  These
protections prevent clients from directly connecting to servers---at least
initially---and all middleboxes succeed at doing so, even if the protection is
identifiable.  However, with the use of more than one source IP address, an
adversary can bypass connection shunning and dynamic blocking and still solicit
\sa{}s from the end-host, albeit rate-limited by the number of scanner addresses.

Beyond actively preventing DDoS attacks and some scanning, each protection
inadvertently slows down the discovery of new services through Internet scanning
and can slow down the spread of malware. Dynamic blocking (completing the
handshake without acknowledging data) is the most effective at doing so. The
technique slows scans by up to 55~times as in the case of host discovery on
27017/MongoDB (Section~\ref{sub:l4_heuristic}), by forcing the scanner to
timeout upon not receiving an \ack for each scanned host. Though zero window
\sa{s} also cause a scanner to eventually timeout, zero-sized windows are easy
to filter.  Immediately closing the connection after the handshake causes only a
negligible slowdown, bounded only by the time it takes to complete a handshake
(about 100\,ms). Connection shunning is the least effective at slowing down
stateless scanners but slows down stateful scanners at the same rate as dynamic
blocking.

\subsection{Summary}
\label{sub:sec:sec2_summary}

Our results establish that \sa{s} are a poor indicator for the presence of a
service. In the worst case, \sa{s} overestimate the hosts that acknowledge data
by 533\% on port~11211 (memcached). We also discover that an average 16\% of services on popular ports and
40\% of services on unpopular ports fail to acknowledge data, which is a likely indicator for the presence of a middlebox protection. We
investigate why hosts that appear to fully speak TCP do not always complete L7
handshakes in the next section.

\begin{figure}[h]
  \includegraphics[width=\linewidth]{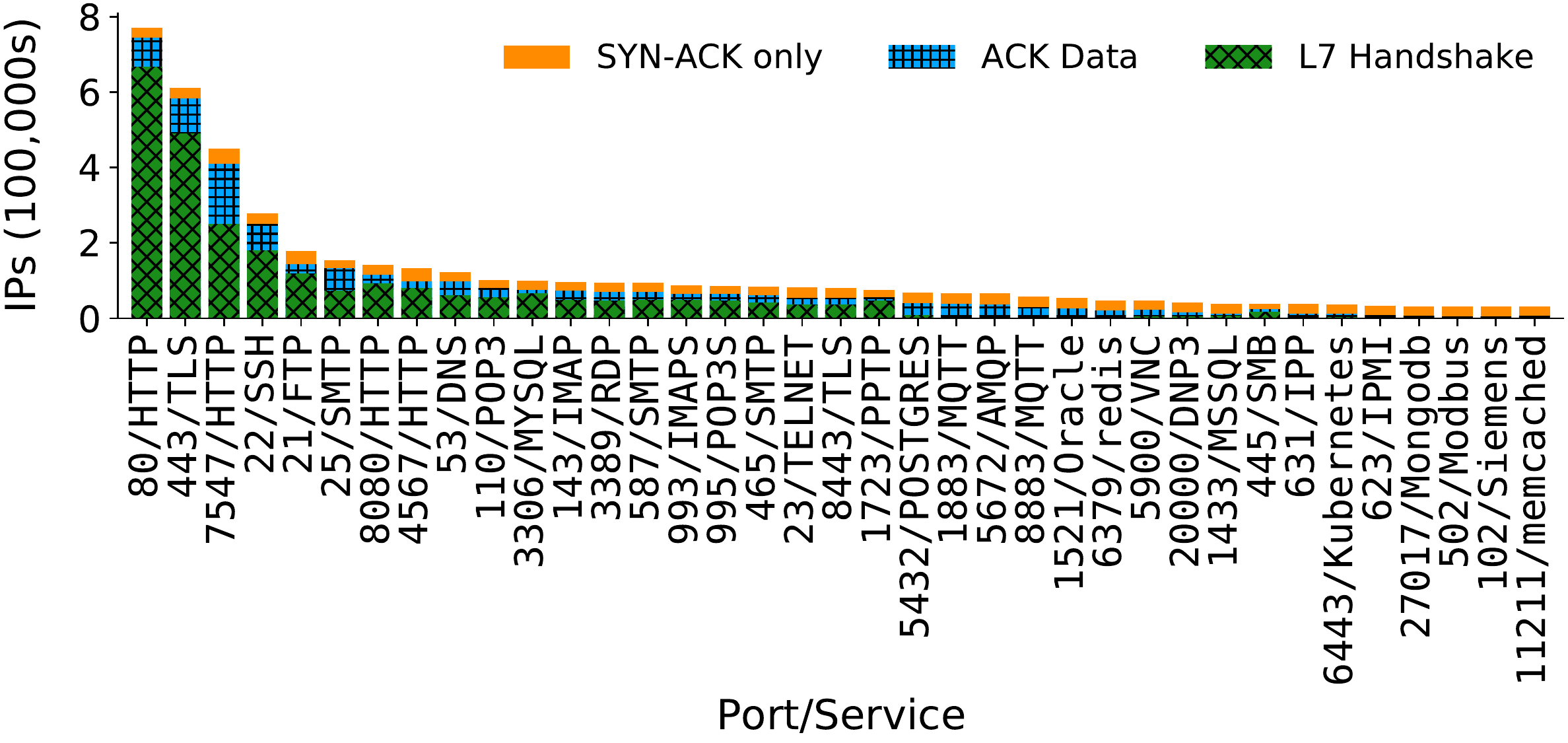}

  \caption{\textbf{SYN-ACK vs.\ Ack.\ Data vs.\ L7 Handshake}---% The number of
	There are up to three orders of magnitude fewer IPs that acknowledge data
	than respond with a \sa packet.
	} \label{fig:sa_v_accept}
\end{figure}

\section{Application-Layer Service Deployment}
\label{sec:l7}

In the last section, we investigated L4-responsive services that do not appear
to speak any L7 service and are artifacts of DoS and scanning protections.
After excluding the 28\% of pseudo-services, we discover 27\% of services on
popular ports and 63\% services on unpopular ports that acknowledge data do not
run the expected application-layer protocol (Figure~\ref{fig:sa_v_accept}). In
this section, we analyze services that complete unexpected application-layer
handshakes or acknowledge data but do not speak any identifiable
application-layer protocol. We show that while IANA-assigned services are
prominent on popular ports, unexpected but identifiable services dominate other
ports. Moreover, assigned ports only host a tiny fraction of the services that
run popular protocols. For example, only 6.4\% of TLS services run on TCP/443.
Services on unexpected ports are commonly hosted by IoT devices and have weaker
security postures, which suggests the need for the security community to study
the services on unassigned ports.

\subsection{Finding Unexpected Services}
\label{sub:sec:unassigned_serv}

To determine the extent to which unexpected services co-reside on ports with
assigned services, we scan 1\% random samples of the IPv4 address space on the
set of ports from Section~\ref{sub:sec:tcp_state} (37~ports with an
expected service and 18~ports without an unexpected service or implemented scanner). For each responsive service, we first
attempt to complete an L7 handshake using the expected protocol, if one exists.
Upon failure, we attempt follow-up handshakes using the 30~protocol
scanners---the total number of \textit{unique} protocol scanners---implemented in ZGrab (Appendix~\ref{app:protos}) with default
parameters.

%\vspace{3pt} \noindent \textbf{Ethical considerations.}
\paragraph[Ethical considerations.] Prior studies have primarily performed
Internet scans that target only expected protocols; to minimize the potential
impact of our experiment, we scan only 1\% of the IPv4 address space.  We
received zero abuse complaints, requests to be blocked from future scans, or
questions from operators from this set of experiments.

\paragraph[Data acknowledging firewalls.] The number of
data- acknowledging services per IP follows a bi-modal distribution: 98\% of IPs
serve fewer than four unidentifiable services and 2\% of IPs host unidentifiable
services on over 60K~ports. About 75\% of all unidentifiable services on
unpopular ports are hosted by IPs with unidentifiable services on nearly every
port (``Unknown Service - across ports'' in Figure~\ref{fig:hist_services}).
Hosts have unidentifiable services on \textit{most} but not \textit{all} ports because some
networks drop all traffic to security-sensitive ports. For example, out of the
top 50~networks that send back the most \sa responses across all ports, 28\%
drop all traffic to port 445 (SMB) and 10\% drop port 23 (Telnet).  Hosts with
unidentifiable services on nearly every port are concentrated in a small number
of networks; five ASes belonging to the Canadian government (74, 25689, 818,
2680, and 806) account for 77\% of all IPs that host unidentifiable services on
nearly every port.

We trace this behavior to the F5 Big-IP Firewall based on a \rst
fingerprint~\cite{f5blogrst} that contains the words ``BIG-IP System.'' An F5
DevCentral blog post~\cite{f5blog} speculates that IPs respond on every port due
to the accidental use of a wildcard when configuring the firewall or an overload
of the firewall's \syn-cookie cache. We identify and exclude these hosts, to
avoid biasing our analysis, by checking whether hosts acknowledge data on five
random ephemeral ports, which effectively filters out 99.9\% of such hosts.
Nonetheless, an average of 10\% of popular and 25\% of unpopular services remain
unidentifiable (i.e., do not respond to any of the 30 handshakes) after
filtering.
\looseness=-1

\subsection{Characterizing Unexpected Services}
\label{sub:sec:unexpected services}

\begin{figure*}[h]
   \centerline{\includegraphics[width=\textwidth]{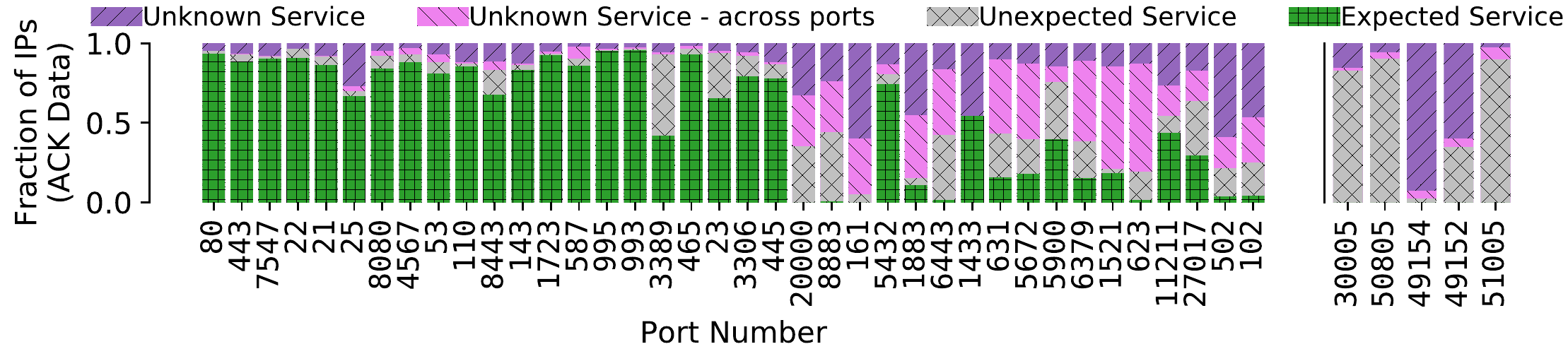}}
  \caption{\textbf{Distribution of Types of Services}---%
	A smaller fraction of services run the assigned protocol on less popular
	ports. For example, only 4\% of services on TCP/102 speak the assigned S7 protocol. The
	fraction of services that can be identified on unassigned ports (on the right hand side) varies widely.
}
  \label{fig:hist_services}
\end{figure*}

After filtering out hosts with unknown services on nearly all ports, we investigate
unexpected services on assigned ports and services on ports without any assigned
service. We summarize our results in Figure~\ref{fig:hist_services} and describe
them here.

\begin{figure}[h]
   \centerline{\includegraphics[width=\linewidth]{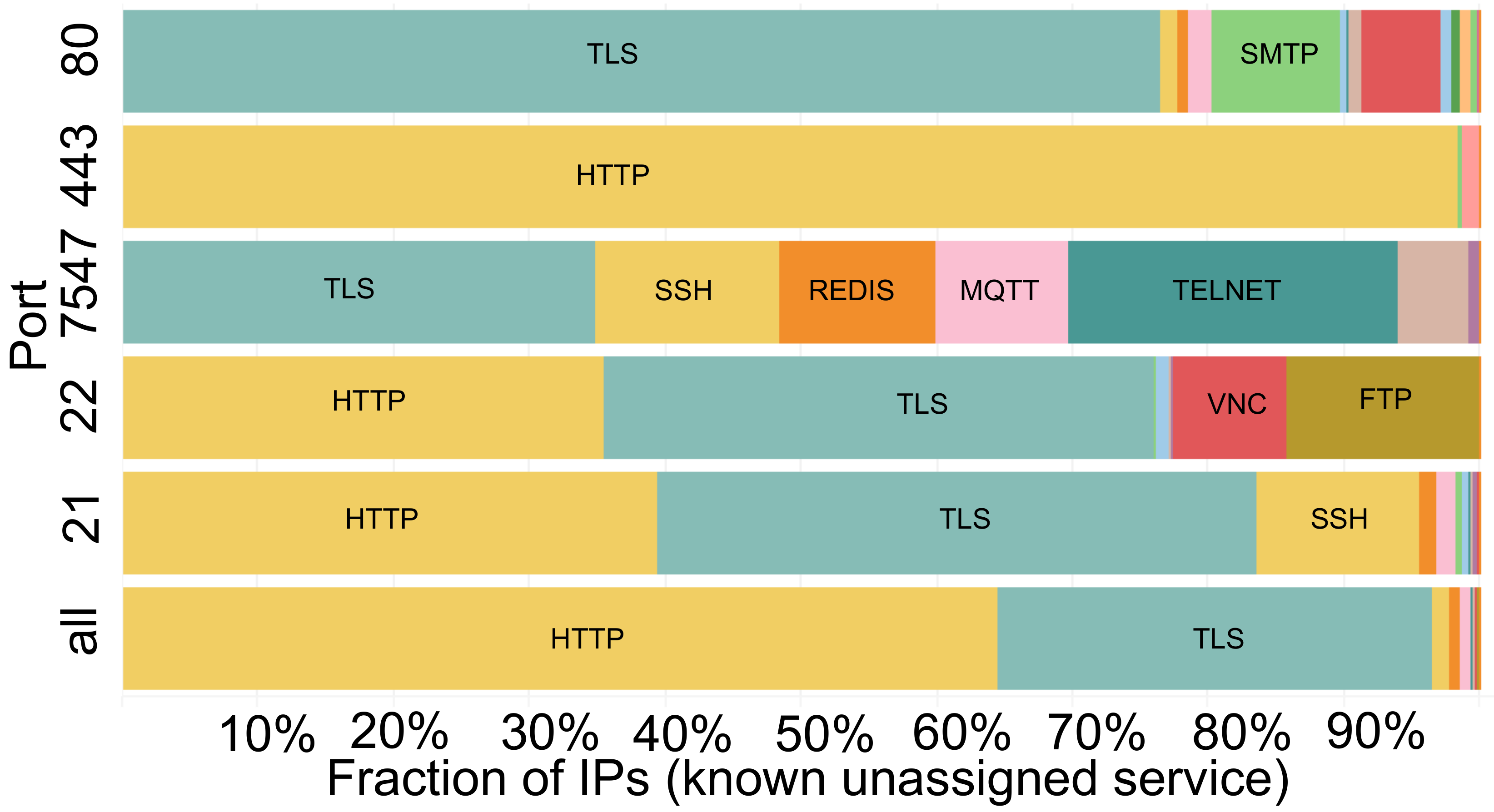}}
  \caption{\textbf{Distribution of Unexpected Services}---%
  HTTP and TLS are the most popular unexpected services, with 65\% of unexpected services speaking HTTP and 30\% speaking TLS.
  }
  \label{fig:80_annot}
\end{figure}

\begin{figure}[h]
   \centerline{\includegraphics[width=\linewidth]{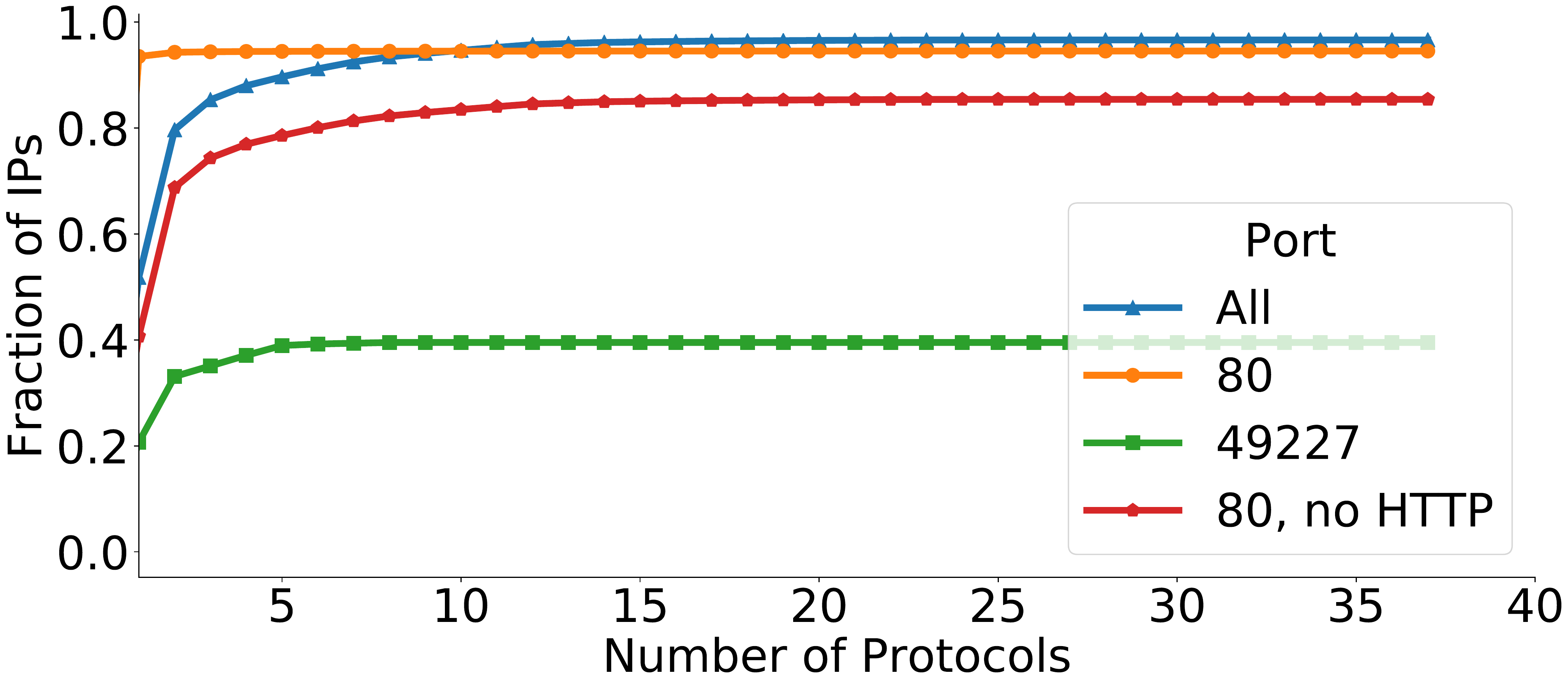}}
  \caption{\textbf{Protocol Coverage Convergence}---The marginal gain of
	scanning additional
  protocols is negligible beyond the top 10~protocols. Still, for most ephemeral ports (e.g., port~49227) the majority of services remain unknown.
  }
  \label{fig:protocol_cdf}
\end{figure}

\paragraph[Unexpected services.] Services on popular ports
typically run the expected protocol: 93\% of hosts that acknowledge data on
port~80 respond to an HTTP GET request and 89\% on port~443 complete an HTTPS
handshake (Figure~\ref{fig:hist_services}). Only 1.6\% of the services on
port~80 and 4.25\% of services on port~443 respond to one of the other
30~unqiue handshakes. The majority (75\%) of unexpected services on port~80 are
TLS-based and nearly all on port~443 are HTTP-based (Figure~\ref{fig:80_annot}).
This implies that operator recommendations to run services on ports~80 or~443 to bypass firewall
restrictions~\cite{firewallbypass} are not widespread.
As ports decrease in popularity, the
fraction of IPs that speak the expected service approaches zero.  For example,
on port~623, only 1\% of services that acknowledge data speak IPMI and 18.9\%
speak other identifiable protocols. Consequently, the number of additionally
identifiable services diminishes after the first few protocols and appears to
converge at 96\% (Figure~\ref{fig:protocol_cdf}). Each port contains its own
long-tail of unexpected services, but for many ports, this number plateaus
quickly---just not at 100\%.

The number of identifiable services on ports without an assigned service varies
between 2--97\% based on port. Among random ephemeral ports, our 30~handshakes
identify the protocol for an average 21\% of services that acknowledge data and
an average of 10~unique protocols per port.  Across all scanned ports, nearly
65\% of unexpected, but identifiable, services speak HTTP and 30\% speak TLS\@.
IoT devices are a prominent culprit behind unexpected services; unexpected TLS
services are 5~times more likely and unexpected SSH 2~times more likely to
belong to an IoT device than 443/TLS and 22/SSH services, respectively. We also
find evidence of operators attempting to hide services. For example, 70\% of
hosts serving TLS on the random ephemeral ports~49227, 47808, and 49152 are
issued certificates by BBIN International Limited, a Philippine offshore online
gambling platform~\cite{bbin}. We further detail the types of services hosted on
unassigned ports in Sections~\ref{sub:sec:eval_security}.

\begin{figure}[h]
   \centerline{\includegraphics[width=\linewidth]{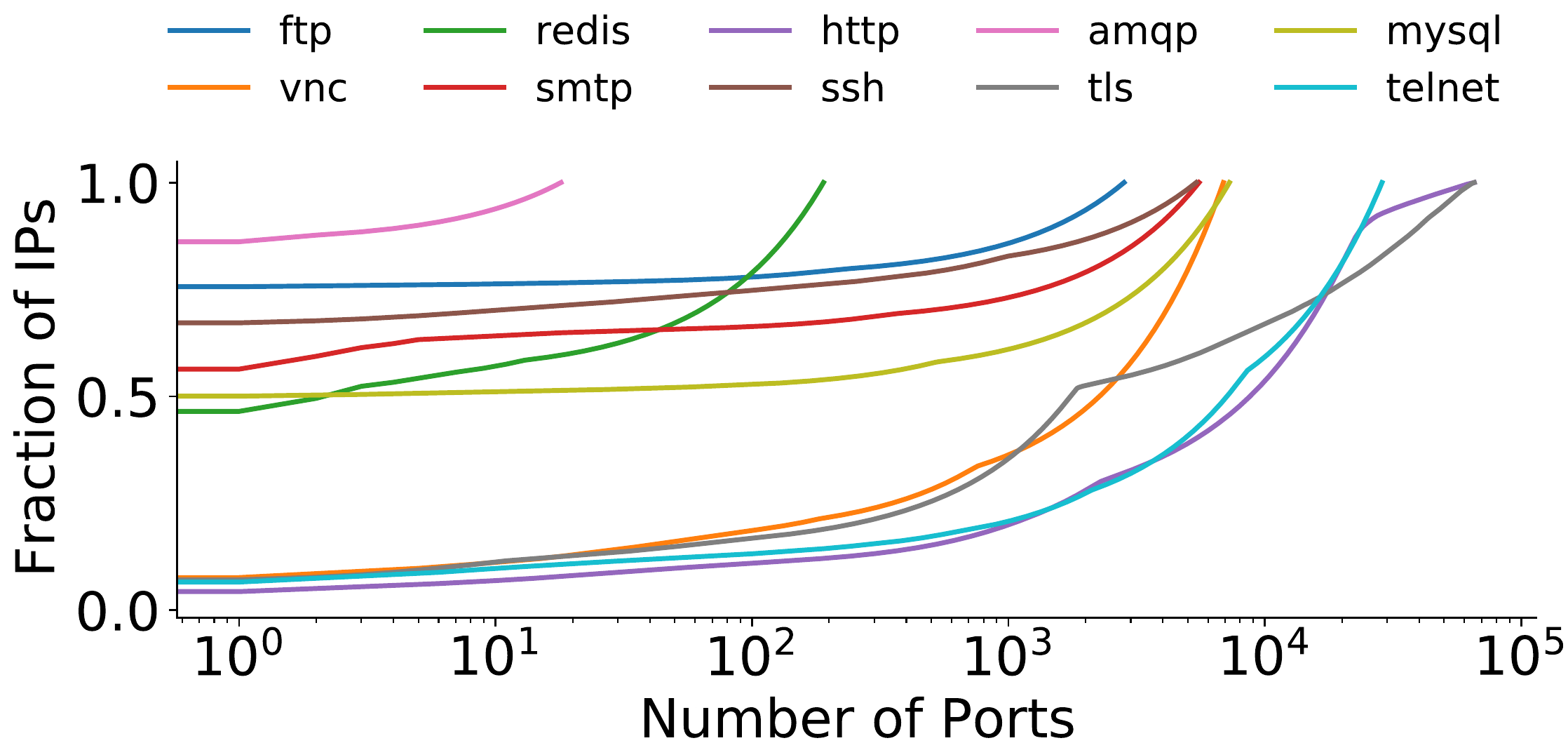}}
  \caption{\textbf{Protocol Coverage Across Ports}---Only 3.0\% of HTTP services
	are served on port~80. Researchers must scan 25K~ports to achieve 90\%
	coverage of HTTP services. On the other hand, 83.1\% of AMQP
	services are on port~5672.
}
  \label{fig:port_protocol}
\end{figure}

\paragraph[Long tail of ports by protocol.] Our results suggest that protocols
run on many additional ports beyond their primary IANA-assigned port. To
quantify how many ports researchers need to scan to achieve coverage of a
protocol, we conduct a new scan targeting 0.1\% of the IPv4 address space on
10~popular protocols on all 65,535~ports and compute the fraction of hosts
running a given service across multiple ports (Figure~\ref{fig:port_protocol}).
We find that port~80 contains only 3.0\% of hosts running HTTP; another 1.2\% of
HTTP hosts run on port 7547 and 0.7\% on port 30005. To cover approximately 90\%
of HTTP, one must scan 25,000~ports.  Only 5.5\% of Telnet resides on TCP/23,
with the assigned alternative port TCP/2323 being only the 10th most popular;
other unexpected ports dominate the top-10 ports with the most Telnet services
(Table~\ref{table:telnet_coverage}).  Previous work tracking botnet
behavior~\cite{mirai,kumar2019early} has primarily studied assigned Telnet ports
(i.e., 23, 2323); our findings imply that the attack surface and number of
potentially vulnerable devices is potentially over 15~times worse than
previously shown.
\looseness=-1

Some protocols are still relatively clustered around their assigned ports. For
example, 83.1\% of all AMQP is on port~5672 and an additional 3.1\% is on
port~5673. HTTP and TLS are the only two protocols which appear on every port in
our 0.1\% IPv4 scan. The set of most popular ports also varies per protocol and
is often not correlated with the popularity of ports that send data (i.e.,
across all protocols), as most services are drowned out by the overwhelming
popularity of HTTP and TLS\@. For example, 7 of the top~10 ports most likely to
host Telnet are ranked above 12,000 in overall popularity.  As a result, when
choosing which popular ports to study for a specific protocol, we recommend
researchers conduct a lightweight sub-sampled scan across all ports.

\begin{table}[h]
%\vspace{-5pt}
\centering
\small
\begin{tabular}{lrll}
\toprule
Port & Hosts & Top AS  & \% of Hosts  \\
& &  & in Top AS  \\ \midrule
23 &	2,606 &	Telecom Argentina (10318) & 8.7\%	\\
5523  & 521 &	Claro S.A (28573)& 87\%	\\
9002 &	396 &	Fastweb Italia (12874)& 4\%	\\
6002 &	232 &	Fastweb Italia (12874)& 6\%	\\
8000 &	158 &	Powercomm KR (17858)&  89\%	\\

\bottomrule
\end{tabular}
	\caption{\textbf{Top~5 Ports Hosting Telnet}---%
	While Telnet is most often seen on its assigned port (TCP/23), the majority of
	Telnet services are served on unassigned ports. Unexpected Telnet devices
	are sometimes spread across a large number of ASes (e.g., port~9002) and are therefore likely not due to a single operator decision.
	 }
	 \label{table:telnet_coverage}
\end{table}

\subsection{Security of Unexpected Services}
\label{sub:sec:eval_security}

Services on unexpected ports are more likely to be insecure than services on
assigned ports. We use the results from our experiment in
Section~\ref{sub:sec:unassigned_serv} (scanning 30~protocols on 55~ports) to
show four examples of how unexpected services affect the results of previous and
future security studies.

\paragraph[IoT devices.] IoT devices are frequent targets due to their
consistently weak security designs~\cite{yang17, lin17, fruscati18}. While
passive measurement has shown that a significant number of IoT devices inhabit
non-standard ports~\cite{kumar19}, active measurement of IoT devices has largely
studied only standard ports~\cite{samarasinghe2019another, yu2015handling,
feng2018acquisitional, agarwal2009netprints, cui2010quantitative, pour2020data}.
By manually identifying server certificates belonging to an IoT manufacturer, we
find IoT interfaces on unexpected ports are widespread; 50\% of TLS server
certificates on unexpected ports belong to IoT devices and unexpected TLS is
5~times more likely to belong to an IoT device than on port~443. For example,
35\% of 8000/TLS are icctv devices (i.e., surveillance cameras) in Korea Telecom
and 38\% of 80/TLS are Huawei network nodes spread across 1\% of all
international networks.  About 5\% of TLS on port~8443 belongs to Android TVs in
Korean networks and at least 20\% belongs to routers. Unassigned ports also
contain more TCP/UPnP devices. For example, there are 12~times more TCP/UPnP
devices on port~49152 (primarily in Latin America and Asian Telecoms) and
2~times as many on ports 58000 and 30005 than on port~80.
%We also note that even though ports~30005, 58005 and 49152 are amongst the
%15~ports with the most responsive services, Shodan~\cite{shodan} does not
%currently scan these ports.

\paragraph[Vulnerable TLS.] TLS services on unassigned ports are
1.17~times more likely to have a certificate with a known private key than on assigned ports. When scanning unassigned ports, we find over twice as many certificates have a known private key than reported
in prior work~\cite{heninger2012mining,hastings2016weak}. For example, 40.2\% of TLS
hosts on port 8081 are DOCSIS~3.1 Wireless Gateways in Telecom Argentina
(AS\,10481 and 10318) using the same OpenSSL Test Certificate with a known
private key and 39\% of TLS hosts on port 58000 are Qno wireless devices with
the same self-signed certificate with a known private key.  Across 23\% of
scanned ports, public keys are more likely---up to 1.7~times more---to be shared
than those on port~443 (e.g., 80/TLS is 1.5~times more likely). Nonetheless,
previous work studying cryptographic keys on the
Internet~\cite{heninger2012mining, hastings2016weak, durumeric2013zmap} has
limited analysis to 443/HTTPS, 22/SSH, 995/POP3S, 993/IMAPS, and 25/SMTPS.

\paragraph[Login pages.] Over half of unexpected ports scanned host a higher fraction of
public-facing login pages (i.e., HTML containing a
login, username, or password field) than
80/HTTP and 443/HTTPS. Though the total number of
HTTP login pages is greatest on port 80, a page on 8080/HTTP is 2.4~times more
likely to be a login page, thus offering an additional 25\% of such pages
compared to port 80.  Furthermore, all the aforementioned IoT devices (e.g.,
icctv, routers) hosting TLS also serve a login HTTPS page on their respective
ports.

\paragraph[SSH hygiene.] Unexpected ports hosting SSH are
15\% more likely to allow non-public key authentication methods (e.g., password, host-based, challenge-response)
than 22/SSH and 2.4~times less likely to be using \textit{only} public key
authentication (11\% vs. 26\%). 60\% of scanned ports are on average 2~times
more likely  (9\% vs.\ 18\%) to be running a software implementation of SSH that is likely to be
on an IoT device (e.g., Dropbear, Cisco, Huawei).

\subsection{Summary and Implications}

Most services that acknowledge data on popular IANA-assigned ports run the
expected L7 protocol, but this drops to nearly zero for less popular protocols
with assigned ports. The majority of services that speak popular protocols
(e.g., TLS, Telnet, HTTP) are spread across all 65K~ports rather than on their
assigned port(s). For example, only 3\% of HTTP services listen on port~80.
Many of the services listening on random ports belong to IoT devices and/or have
a weak security posture, and it behooves the security community to consider
these services when quantifying risk.

\section{Efficiently Identifying Services}
\label{sec:optimizations}

L7 scanning is more challenging when there is no assigned protocol for a port or
when the expected L7 handshake fails. Though Section~\ref{sub:sec:eval_security}
demonstrates the importance of scanning for unexpected services, the naive
method we used tests 30~unique L7 handshakes and is too intrusive and slow for
large-scale experiments. In this section, we explore how to most efficiently
detect unexpected L7 services. Encouragingly,
only five handshake messages are needed to uncover 99\% of unexpected services running
identifiable protocols.

\subsection{Protocol Discovery}
\label{sub:sec:fingerprinting}

We investigate two directions for accelerating protocol discovery: (1) methods
that trigger protocol-identifying responses on a large number of protocols and
(2) attempting handshakes in an order that optimizes for efficient service
discovery.

\paragraph[Wait and fingerprint.] The most efficient first
step for detecting the protocol on a port is to simply wait to send any
handshake message and to see what the server sends first. A total of 8 of the 30~protocols
implemented in ZGrab---POP3, IMAP, MySQL, FTP, VNC, SSH, Telnet, and
SMTP---are ``server-first'' protocols: after a TCP handshake concludes, the
server will send a banner to the client, which allows the client to parse and
identify the actual service.
For example, 99.99\% of hosts which complete an SSH handshake have the keyword \texttt{ssh} in the SSH banner, 90\% of SMTP
banners contains \texttt{smtp}, 72\% of Telnet contains \texttt{login} or
\texttt{user}, and 100\% of VNC responses contain \texttt{RFB}.  We are able to
identify banner signatures for all implemented binary and ASCII-based protocols.

\begin{figure}[h]
   \centerline{\includegraphics[width=\linewidth]{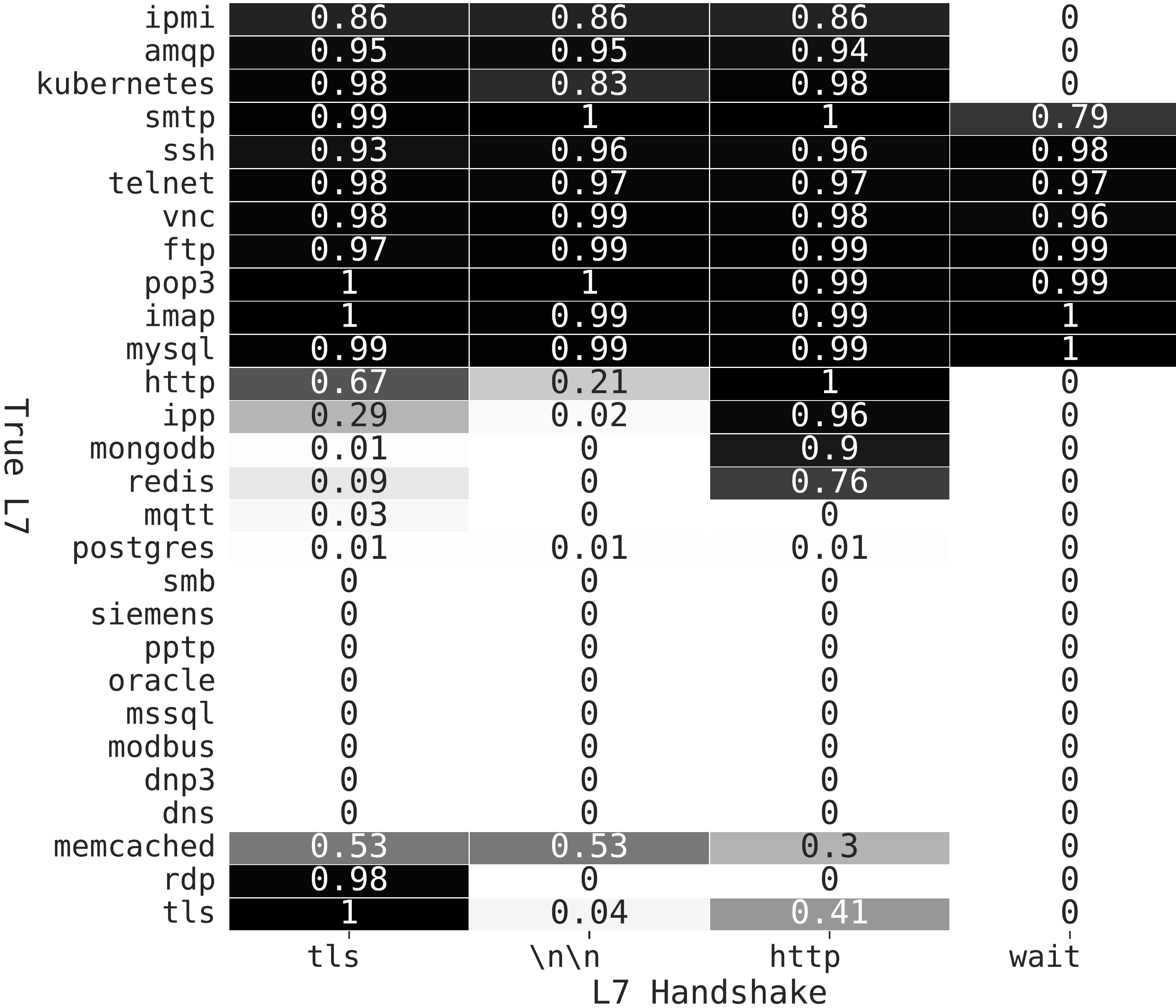}}
  \caption{\textbf{Scanning L7 With Different Handshakes}---%
  Sending an HTTP handshake (i.e., a GET Request) prompts the most number of services to send back data. The data can then be used to fingerprint the actual service running.
  }
  \label{fig:heatmap_cov}
\end{figure}

We also find that many protocols respond to incorrect handshake messages,
including HTTP and TLS.
Through 1\% scans of the IPv4 space, we find that 16 of 30~protocols respond to an HTTP GET request or two newline characters for at
least 50\% of public services that speak the protocol
(Figure~\ref{fig:heatmap_cov}). In general, most services that respond to the
wrong handshake respond to both a GET request and TLS Client Hello, but
MongoDB, and Redis do not send data in response to a TLS handshake. Though sending two newline characters is
protocol-compliant for many ASCII protocols, doing so discovers fewer services than TLS and HTTP\@. We discover a similar phenomenon
when sending 50~newline characters, thereby implying that the contents of the newline
message---rather than the length---causes the lack of
responses.

A total of 75\% of binary (i.e., non-ASCII) services, including MQTT, Postgres, PPTP, Oracle
DB, Microsoft SQL, Siemens S7, DNS, and SMB, send no data back unless we scan
with their specific protocol. We note that our selection of tested protocols are biased
towards ASCII protocols, and that it is likely that many binary protocols do not
respond to these handshake messages. However, as discussed in
Section~\ref{sub:sec:unexpected services}, the long tail of binary protocols on the Internet are less
spread out across a large number of ports compared to common protocols like
HTTP\@.

\begin{table}[t]
\centering
\small
\begin{tabular}{ccccc}
\toprule
Scan
&
\multicolumn{2}{c}{IANA-Assigned Ports}
&
\multicolumn{2}{c}{Ephemeral Ports}  \\ \cmidrule(r){2-3}\cmidrule(l){4-5}

Order & Protocol &  $\Delta$ Coverage & Protocol &  $\Delta$ Coverage   \\
	\midrule 1 & \emph{wait} & 51.3\% & \emph{wait} & 66.3\%  \\ 2 & TLS &
	29.0\% & HTTP & 17.1\%  \\ 3 & HTTP  & 13.6\% & TLS & 15.9\% \\ 4 & DNS &
	3.4\% & Oracle DB & 0.23\% \\ 5 & PPTP & 1.8\% & PPTP & 0.14\% \\
\bottomrule \end{tabular}
	\caption{\textbf{Optimal Handshake Order}---% Scanning
	For IANA-assigned ports, waiting and then sending a TLS Client Hello
discovers 80.3\% of unexpected services. Five handshakes can identify over 99\% of identifiable unexpected services.}
\label{table:l7_coverage}
\end{table}

\paragraph[Optimal handshake order.]
We compute the optimal order of L7
handshakes that maximize the chances of identifying the service running on a
port using a greedy approach across two sets of ports: (1)~all IANA-assigned
ports and (2)~five random ephemeral ports (62220,
53194, 49227, 47808, and 65535). Of the 30~protocols with ZGrab scanners that
we can identify, we find that five handshake messages elicit responses from
over 99\% of identifiable unexpected services on both sets of ports. We show the top-five L7
handshakes that discover the most unexpected services for the two sets of ports,
excluding the expected services in Table~\ref{table:l7_coverage}. Across both
IANA-assigned and ephemeral ports, merely opening a connection to the client
(i.e., waiting) can immediately fingerprint more than half of unexpected
services. For IANA-assigned ports, waiting and then sending a TLS Client Hello
discovers 80.3\% of unexpected services. For ephemeral ports, waiting and HTTP
discover 83.4\% of services. It is not surprising that DNS and PPTP provide the
4th and 5th most additional coverage for IANA-assigned ports, as these are
relatively popular protocols that do not answer to other handshakes (e.g., HTTP~GET).

\subsection{Impact of L7 Filtering}

One reason that we may not be able to identify all services is that even if our
protocol guess is correct, our selected handshake parameters might be rejected.
For example, in SNMP, servers may reject requests that do not specify the
correct community string in the first packet by first acknowledging the data,
but then sending a TCP \rst. To estimate whether L7 filtering decisions cause a
service to not send \textit{any} data back to the client, thereby hindering
fingerprinting efforts, we run two sets of scans, each with different handshake
options, for each of the following ports and protocols: 8081/HTTP, 443/TLS, and
1723/PPTP.

For HTTP, in one scan we send a GET request and in another we specify the
OPTIONS request.  For TLS, in one scan we advertise the insecure cipher suite
TLS\_RSA\_EXPORT\_WITH\_RC4\_40\_MD5 and in the other we advertise modern Chrome
cipher suites. For PPTP, in one scan the first message is crafted to contain the
specified ``Magic Cookie'' value (a specific constant used to synchronize the
TCP datastream) according to RFC~2637~\cite{pptp}, \texttt{0x1A2B3C4D}, and in
another we specify the Magic Cookie to be \texttt{0x11111111}. RFC~2637 states
that ``Loss of synchronization must result in immediate closing of the control
connection's TCP session;'' we thus expect that fewer IPs will send data to the
client if the magic cookie is incorrect and use this as a ``control''
experiment.

\begin{table}[h]
	\centering
	\small
	\begin{tabular}{lll}
		\toprule Port (Service) & Handshake Option & IPs that send data\\ \midrule
		& Only GET Request & 27\% \\
		8081 (HTTP) & Only OPTIONS Request & 7.3\% \\
		& Both  & 65.7\% \\
		\midrule
		& Only Good Cookie & 67.1\% \\ 1723
	(PPTP) & Only Bad Cookie & 0.001\% \\ & Both  & 32.8\% \\ \midrule
		& Only Secure Cipher & 2.65\% \\
		443 (TLS) & Only Insecure Cipher &0.05\% \\
		& Both & 97.3\% \\
%\midrule & Only Secure Cipher & 1.8\% \\ Port 6443 (TLS) & Only insecure Cipher
	%&0.2\% \\ & Both  & 97.9\% \\
\bottomrule
\end{tabular}
\caption{\textbf{Impact of Handshake Options}---%
	Handshake parameters influence the services that send back identifiable
	data. For example, an HTTP OPTIONS request on port~8081 results in 7.3\%
	more IPs to respond with data than an HTTP GET request. 65.7\% of IPs will
	respond to both types of requests on port~8081.
}
\label{table:options}
\end{table}

An HTTP OPTIONS request discovers an additional 7.3\% IPs that speak HTTP compared to a
GET request on port~8081. Responsive IPs will acknowledge data and close the
connection after receiving a GET request, hindering a scanner's ability to
fingerprint the service as HTTP\@. However, by sending an OPTIONS request, 72\%
of IPs will respond with a 501 status (method not implemented) and 17\% will
respond with a 405 status (method not allowed), thereby confirming they do speak
HTTP\@. IPs that exclusively respond to an OPTIONS request are not
constrained to a particular network and are present across 5.3\% of ASes.  The
discrepancy is less pronounced on port~80 where only 0.02\% of IPs will respond
to an OPTIONS request but not GET and only 1.1\% of IPs will respond to GET but
not an OPTIONS request.

For TLS, per RFC\,8446~\cite{rfc8446}, a handshake failure should generate an
error message and notify the application before closing the connection.
However, 2.65\% of IPs will simply close the connection without any
application-layer error when an incompatible cipher is given.  As expected for
PPTP, specifying an incorrect magic cookie results in 67.1\% of IPs failing to
respond (Table~\ref{table:options}). Hosts practicing their own Layer 7
filtering depending upon certain handshake options---and thereby not sending any
data to the client---presents an unavoidable challenge for any L7 scanner to
guess the perfect parameters to speak the appropriate Layer~7 with every single
host. In Figure~\ref{fig:hist_services}, we estimate all unknown services to be
due to not having the expected handshake options.

\subsection{Consequences of Handshake Order}
\label{sub:sec:consequences_handshake_order}

Similar to how handshake options might prevent a server from responding, trying
repeated incorrect handshakes prior to the correct one might also prevent the
identification of services. We evaluate whether hosts filter or refuse
connections after receiving incorrect L7 messages by (1) sending successive HTTP~GET
and TLS Client Hello messages to all IANA-assigned ports for 1\% of the IPv4
space and (2) comparing the number of hosts that successfully complete a follow-up
handshake when being sent the expected L7 data to the number of hosts that
successfully complete a follow-up handshake when being sent unexpected L7 data.

\begin{figure}[h]
   \centerline{\includegraphics[width=\linewidth]{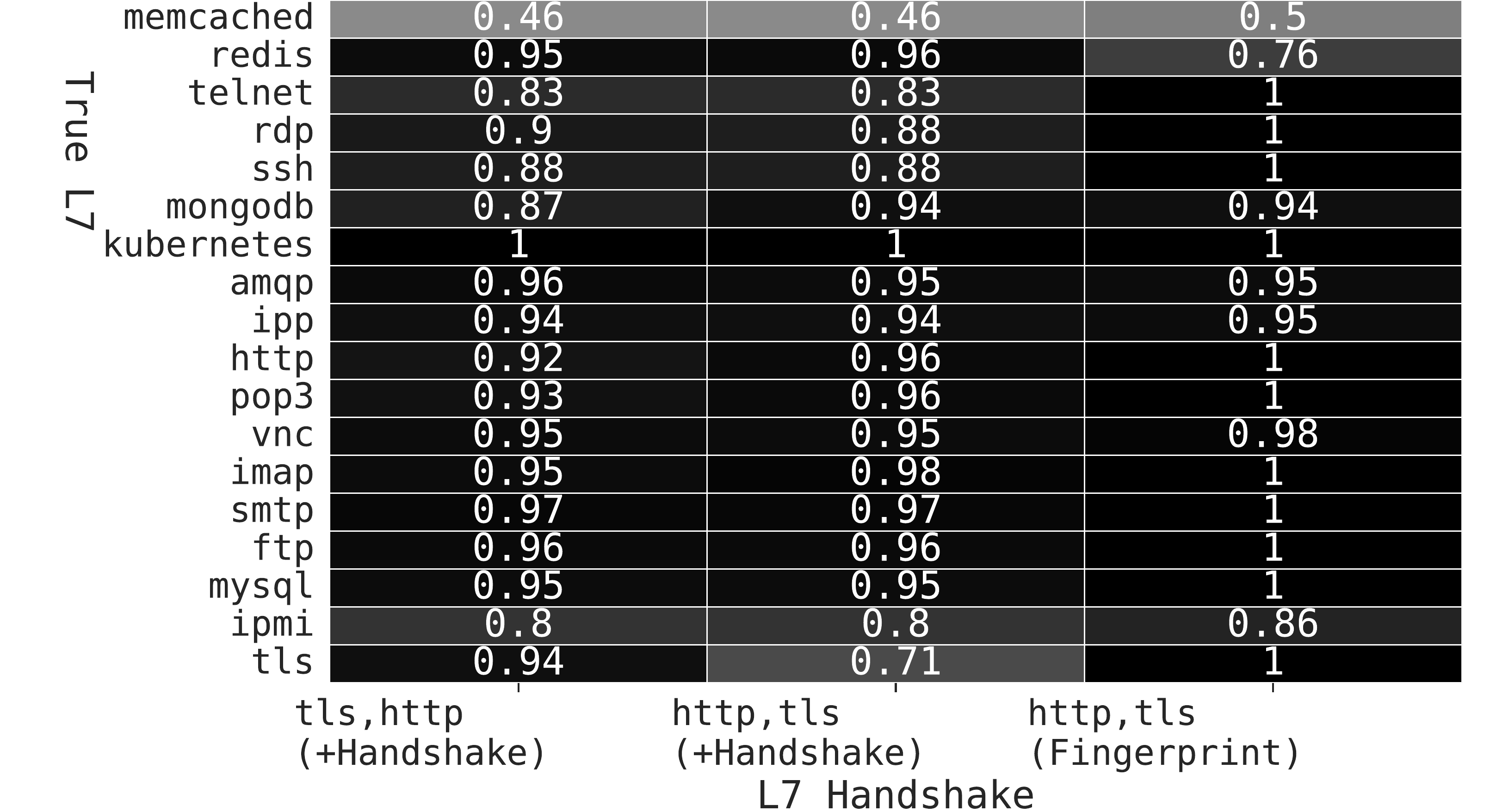}}
  \caption{\textbf{Impact of Sending Incorrect Handshakes}---%
  Sending unexpected data to hosts causes some services to fail the follow-up expected handshake even when fingerprinting was successful. For example, only 71\% of TLS hosts successfully complete a handshake when initially being sent an HTTP handshake message.
  We provide the fraction of total hosts successfully fingerprinted in the third column.
  }
  \label{fig:heatmap_httptls}
\end{figure}

Depending on the protocol, we find that sending unexpected L7 data causes up to
30\% of follow-up handshakes to fail compared to the hosts found when directly
scanning for the protocol (Figure~\ref{fig:heatmap_httptls}). For example,
sending non-Telnet data to Telnet servers causes 17\% to fail a follow-up
handshake; 65\% send a TCP \rst and 35\% do not \sa to a follow up TCP
handshake.  Sending an HTTP GET request to TLS servers causes 29\% of follow-up
TLS handshakes to fail.  We find this behavior to be similar to a Cisco IOS
feature, Login Block, which allows administrators to temporarily block
connections to L7 services after unsuccessful login
attempts~\cite{ciscologinblock}.  Surprisingly, this phenomenon only affects
hosts \emph{after} they send protocol-identifying data---likely because this is
when they first store server-side application-layer state about the connection.
As such, this blocking does not prevent any servers from being fingerprinted. It
only prevents a follow-up handshake after identifying data has been sent back to
the scanner. Failure is generally temporary: 75\% of hosts will successfully
complete the L7 handshake within 5~seconds and 99\% of hosts will take less than
2~minutes. Nonetheless, waiting between fingerprinting and completing the
follow-up handshake can reduce this filtering effect.

\subsection{Summary and Implications}

One fundamental limitation of L7 scanning is that services may require specific  handshake options to respond.	Nonetheless, our results indicate that the vast majority of
identifiable Internet services can be easily identified during scans. Many hosts
respond to the ``wrong'' L7 handshake and send data that help fingerprint the
service: 16~of 30 protocols can be detected with a single HTTP GET request and
99\% of unexpected services can be identified with five handshakes.
We use these optimizations to build a scanner (LZR) dedicated to accurate and efficient unexpected service discovery.

\section{\lmap: A System for Identifying Services}
\label{sub:l4_heuristic}

In this section, we introduce \lmap, a scanner that accurately and efficiently
identifies Internet services based on the lessons learned from
Sections~\ref{sec:l4}--\ref{sec:optimizations}. \lmap can be used with ZMap to
quickly identify protocols running on a port, or as a shim between ZMap and an
application-layer scanner like ZGrab, to instruct the scanner what follow-up
handshake to perform. \lmap's novelty and performance gain is primarily due to
its ``fail-fast'' approach to scanning and ``fingerprint everything'' approach
to identifying protocols. It builds on two main ideas:

\begin{figure*}[h]
   \centerline{\includegraphics[width=\textwidth]{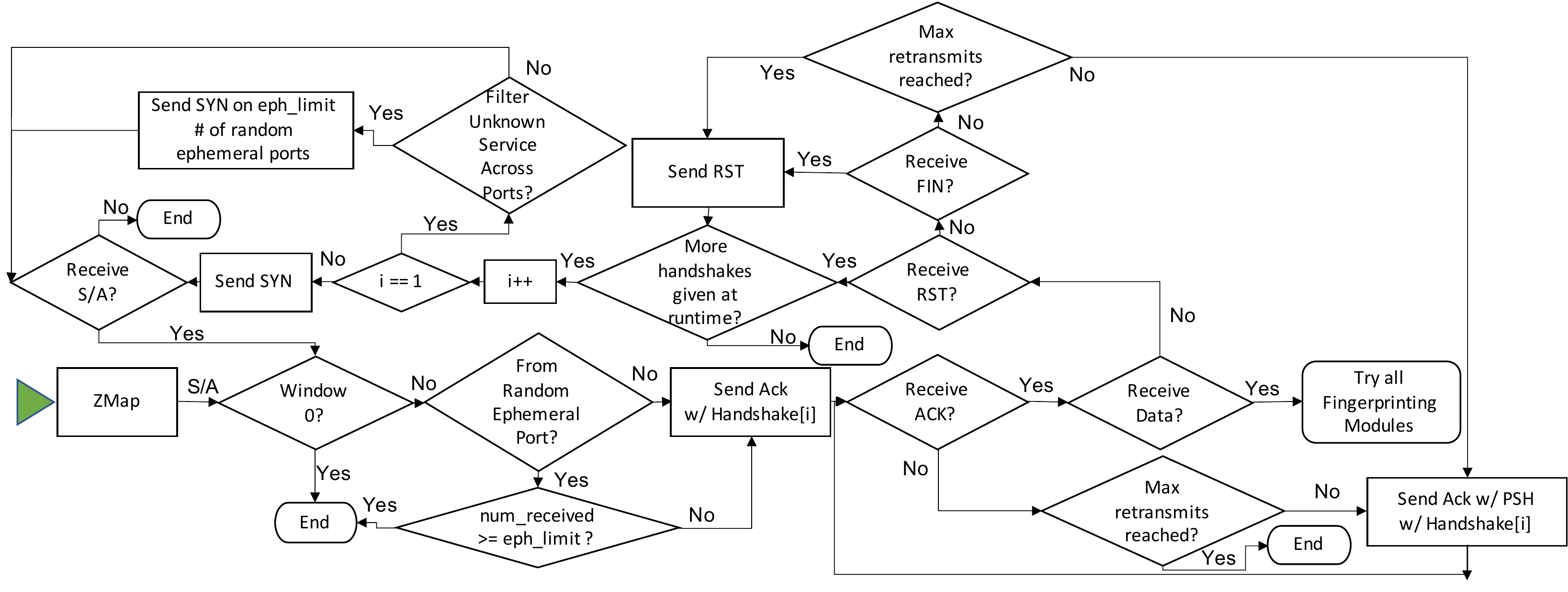}}

  \caption{\textbf{LZR Algorithm}---%
	\lmap efficiently identifies real Internet services by sending
	application-layer data with the \ack of a TCP handshake to filter out
	non-acknowledging hosts and fingerprint the responding protocol.
   }
  \label{fig:lzr_flow}
\end{figure*}

\paragraph[Ignore non-acknowledging hosts.] About 40\% of services that
send a \synack never acknowledge data. None of these services can complete an L7
handshake and can be safely ignored during Internet scans. Quickly
identifying and ignoring these services can significantly reduce costs because
non-acknowledging services force stateful scanners to open an OS socket and wait
for the full timeout period to elapse, which typically takes much longer than
completing a normal handshake. Non-acknowledging hosts can be filtered out by
sending a single packet---an \ack with data---similar to how ZMap statelessly \syn scans.

\paragraph[Listen more.] Up to 96\% of services per port run unexpected
protocols. In 8~of the 30~protocols we scanned, the server sends data first, and
10~protocols send fingerprint-able data when sent an incorrect L7 handshake.
By always waiting and then fingerprinting invalid server responses, we can identify
up to 16~of the 30~protocols by sending a single packet. A scanner only needs to perform
minimal computation to fingerprint a service: the first packet from a server
identifies the running protocol, which does not require a full TCP/IP stack.

\subsection{Scan Algorithm}

We outline \lmap's logic in Figure~\ref{fig:lzr_flow}.
\lmap accepts a stream of \sa{} packets from ZMap or tuples of (IP, port) to
scan. In the case that \lmap has full connection details from ZMap, \lmap will
start by filtering hosts that send \sa{}s with a zero window. Otherwise, it will
initiate a new connection. For non-zero windows, \lmap will continue the connection by
sending an \ack packet containing the expected protocol's first-packet handshake
data. If \lmap receives any type of data in response from the host, it will
fingerprint the data and close the connection. If a host neither acknowledges
the data nor closes the connection, \lmap re-transmits the data with the \psh
flag (further discussed in Section~\ref{sub:sec:lzr_eval}). If a host does not
acknowledge the data (e.g., never responds or \rst{s} the connection without an acknowledgement), \lmap
fingerprints the host as likely not hosting a real service and does not proceed
with further connection attempts. Otherwise, if a host acknowledges the data but
does not send any data in response (i.e., server is unresponsive or closes the
connection immediately afterwards), \lmap proceeds to close the connection,
start a new connection, and send the next handshake. The process continues until
\lmap identifies the running protocol or runs out of additional handshakes to
try. \lmap can also optionally filter IPs that respond on nearly every port
(Section~\ref{sub:sec:unassigned_serv}) by simultaneously sending \syn packets to a
user-specified number of random ephemeral ports and checking for a \sa.

\subsection{Architecture}

\lmap is written in 3.5K~lines of Go and implements all unique protocols (i.e.,
handshakes) in Appendix~\ref{app:protos}.  Similar to ZMap, \lmap uses
libpcap~\cite{linuxpcap} to send and receive raw Ethernet packets rather than
rely on the OS TCP/IP stack. This allows \lmap to efficiently fingerprint
services because a single socket can be used for the duration of a scan and it
allows \lmap to adopt and continue connections initiated by a stateless scanner
like ZMap. Because \lmap only needs to send and receive a single packet to
fingerprint services, a full TCP stack is not needed.

\lmap takes as input a command-line argument list of protocols to test and a
stream of \sa{}s from ZMap or IP/ports to scan. Internally, a small pool of Go
routines send followup \ack packets containing handshake messages and
fingerprint their responses.  Adding new protocols/handshakes to
\lmap is easy; each handshake implements a \texttt{Handshake} interface that
specifies (1) the data to attach to the \ack packet and (2) what to search for
in a response packet to fingerprint the protocol. Once \lmap receives data to fingerprint, \lmap
first checks if the data matches the fingerprint (specified using the  \texttt{Handshake}
interface) of the protocol being attempted. If not, \lmap checks all the
remaining fingerprints for a match. We note that because ZMap
sends probes using a raw Ethernet socket, \lmap users need to install an
iptables rule to prevent the Linux kernel from sending \rst packets in response
to the \sa{}s it receives. Otherwise, \lmap cannot adopt and continue these
connections. We have released \lmap under the Apache 2.0 license at
\url{https://github.com/stanford-esrg/lzr}.

\subsection{Evaluation}
\label{sub:sec:lzr_eval}

\begin{table*}[h]
\centering
\small
\begin{tabular}{llllllllllll}
\toprule
Port & 80 & 443 & 21 & 23 & 5672 & 5900 & 27017 & 62220 & 80 & 443 & 47808 \\
Protocol(s) & HTTP & TLS & FTP & TEL & AMQP & VNC & Mongo & HTTP & HTTP & TLS & HTTP \\
(Consecutively Scanned) &  &  &  &  & &  &  &  & TLS & HTTP & TLS \\
\midrule
Number of Hosts Found\\
\sa & 62.6M & 51.8M & 14M & 6.4M & 3.5M & 3.5M & 2.4M & 2.6M & 63M & 51.6M & 2.8M  \\
Zero Window & 1.3M & 2.1M & 1.7M & 1M & 899K & 1.2M & 695K & 737K & 1.2M & 1.8M & 742K \\
\rst & 1.7M & 2.3M & 1.1M & 673K & 502K & 730K & 166K & 349K & 1.3M & 1.9M & 31K \\
\ack{s} Data & 55M & 45M & 9.5M & 4.6M & 1.4M & 1.4M & 505K & 628K & 56.3M & 45M & 1.1M \\
L7 Handshake  \\
\quad Expected (\lmap) & 54.66M & 43.7M & 9.2M & 2.71M & 123K & 277K & 73.3K & 38K & 56M & 44.3M & 22.6K \\
\quad Expected (ZGrab) & 54.63M & 43.7M & 9.3M & 2.73M & 123K & 277K & 73.6K & 36K & 56M & 44.4M & 22.7K \\
\quad Unexpected (\lmap) & 238K & 1.3M & 113K & 230K & 260K & 56K & 23K  & 23K & 207K & 758K & 26.5K \\
\quad Unique Unexpected & 18 & 16 & 10 & 10 & 11 & 8 & 14 & 12 & 18 & 16 & 14 \\ \midrule
Speed Up (Time) \\

ZMap/\lmap& 3.3$\times$ &  4.7$\times$ & 2.8$\times$ & 3.9$\times$ &1.9$\times$ & 2$\times$ & 1.6$\times$ &2.7$\times$ & 3.3$\times$ & 6.3$\times$ & 2$\times$   \\

ZMap/\lmap + ZGrab& 1.2$\times$ &  1.1$\times$ & 1.2$\times$ & 2.5$\times$ &1.8$\times$ & 1.9$\times$ & 1.4$\times$ & 2.6$\times$ &1.1$\times$ & 0.95$\times$ &2$\times$   \\

Offline ZMap/LZR + ZGrab& 1.1$\times$ &  1.1$\times$ & 2.1$\times$ & 1.6$\times$ &3.3$\times$ & 4$\times$ & 7$\times$ &5.4$\times$ & 1.1$\times$ & 1.1$\times$ &2.5$\times$   \\

Offline ZMap + \lmap& 4.1$\times$ &  4.1$\times$ & 5$\times$ & 10.7$\times$ & 11.4$\times$ & 13.3$\times$ & 55$\times$ &25.3$\times$ & 5.6$\times$ & 3.4$\times$ & 29$\times$   \\

\midrule

Bandwidth Savings \\

ZMap/\lmap& 60$\%$ &  75$\%$ & 67$\%$ & 78$\%$ & 70$\%$ & 79$\%$ & 66$\%$ & 68$\%$ & 79$\%$ & 84$\%$ & 87$\%$   \\

ZMap/\lmap + ZGrab& -28$\%$ &  -16$\%$ & 3$\%$ & 3$\%$ & 41$\%$ & 46$\%$ & 46$\%$ & 54$\%$ & -16$\%$ & -9$\%$ & 75$\%$   \\

Offline ZMap/LZR + ZGrab& 12$\%$ &  10$\%$ & 36$\%$ & 67$\%$ & 72$\%$ & 68$\%$ & 81$\%$ & 79$\%$ & 5$\%$ & 7$\%$ & 98$\%$   \\

Offline ZMap + \lmap& 49$\%$ &  60$\%$ & 56$\%$ & 69$\%$ & 75$\%$ & 78$\%$ & 87$\%$ & 85$\%$ & 58$\%$ & 68$\%$ & 99$\%$   \\

\bottomrule
\end{tabular}
%\caption{\textbf{\lmap Efficacy}---We show the number of IPs that are filtered out by
%	each step of \lmap.
%}

\caption{\textbf{\lmap Performance}---%
	Filtering for IPs that acknowledge data increases service fingerprinting
	speed by up to 55~times while finding up to 30\% more unexpected services.
	All relative performance numbers are compared to ZGrab and measured at a
	1~Gb/s scanning rate.
}
\label{table:comp_zgrabs}
\end{table*}

We evaluate both the accuracy and performance of \lmap by comparing
protocol-specific ZGrab handshakes with four \lmap configurations. The first two
are the expected use cases:

\vspace{-5pt}
\begin{enumerate}[itemsep=2pt,parsep=2pt]

	\item \textbf{ZMap/\lmap}: We use \lmap with ZMap to identify the
		service running on a port that ZMap finds.

	\item \textbf{ZMap/LZR + ZGrab}: We use \lmap as a shim between ZMap and ZGrab to instruct ZGrab what full L7 handshake to complete for hosts that ZMap
		finds.

\end{enumerate}

\noindent During experiments with these configurations at 1gbE, we find that \lmap is able to filter hosts
\emph{much} faster than ZMap is able to find hosts---especially on ephemeral ports with low hitrates.
ZMap artificially limits how fast \lmap and ZGrab operate. As such, we introduce two additional
metrics that approximate \lmap's performance under the premise of ZMap finding hosts infinitely quickly. This allows us
to compute how quickly \lmap can find hosts as scan speeds increase and how much
time ZGrab can save in an environment where there are many hosts to scan because
the researcher is investigating multiple ports simultaneously.

\vspace{-5pt}
\begin{enumerate}[itemsep=2pt,parsep=2pt]
\setcounter{enumi}{2}

	\item \textbf{Offline ZMap/\lmap + ZGrab}: We perform scans in two phases.
		In the first, we use ZMap and \lmap to identify Internet hosts that
		speak a known protocol and exclude this phase from our benchmarking.
		Then, in a second phase, we allow ZGrab to process services at full
		speed.

	\item \textbf{Offline ZMap + \lmap}: We perform scans in two phases.  In the
		first, we find candidate services with ZMap, and exclude this phase from
		our benchmarking. In the second phase, we benchmark how quickly \lmap
		can fingerprint services operating at full speed.

\end{enumerate}

\noindent We report L4 and L7 behavior breakdown, CPU time, and bandwidth savings
of \lmap from 100\% scans of the IPv4 address space completed during June 2020
in Table~\ref{table:comp_zgrabs}.
We calculate runtime performance using CPU cycles per second for ZGrab and
\lmap as both tools are CPU bound: ZGrab's completion of a full handshake (e.g.,
encryption/decryption for TLS) and \lmap's fingerprinting (e.g., pattern
matching) create the biggest performance bottlenecks for each. When benchmarking
\lmap, we receive complaints from seven different organizations, but
there is no indication that the complaints are the result of a particular \lmap
optimization; we follow-up with all responsive network operators and learn that
the complaints are simply due to the 100\% coverage of the scans.

\paragraph[How many additional services does \lmap find?] One of \lmap's key
features is that it can identify additional services, while filtering out
unresponsive ones by analyzing the response to the data included in the \ack
packet. Using the keyword-fingerprinting strategy, \lmap identifies an average
of 12 additional unique protocols across ports in our experiment by using
\textit{only} the expected 1--2 handshakes; for example, 1.3~million IPs hosting
an additional 16~protocols on port~443 and 238,000~IPs hosting an additional
18~protocols on port~80 are found with just the \textit{single} expected
handshake. Furthermore, LZR finds over 2~times more unexpected than expected
services when sending a single AMQP handshake to 5672/AMQP.  The breakdown of
the unexpected services is, unsurprisingly, nearly identical to the distribution
in Figure~\ref{fig:hist_services} (i.e., HTTP and TLS dominate). Across all
ports in Appendix~\ref{app:protos}, LZR identifies 88\% of all identifiable
services with just a single HTTP handshake message.  The exact signatures LZR
uses for fingerprinting services can be found
at~\url{https://github.com/stanford-esrg/lzr/tree/master/handshakes}.

\paragraph[Does \lmap filter out appropriate hosts?] \lmap does not find a
statistically significantly different set of hosts than scanning with just ZMap
and ZGrab (Table~\ref{table:comp_zgrabs}). The Kolmogorov–Smirnov (KS)
test~\cite{ks_test} finds $p>0.05$, rejecting the hypothesis that the approaches
find a different number of services for all tested ports. We also verify that
sending data with an \ack during the handshake does not produce a statistically
significant difference in the total number of hosts that acknowledge data or the
total number of IPs that send back data across three trials of 1\% IPv4 samples
for 80/HTTP, 443/TLS and 27017/MongoDB. However, we do find that an additional
average of 0.18\% of hosts respond when setting the \psh flag during the
retransmission. Though the addition of the \psh flag causes the follow-up packet
to not qualify as an exact TCP retransmission per RFC~793\cite{tcp}, we confirm
that there is no increase in the number of closed connections when
re-transmitting with a \psh flag compared to an identical retransmission. We do
not set the \psh flag immediately during the handshake as that causes about
0.6\% of IPs to close the connection.
\looseness=-1

\paragraph[How much faster is L7 scanning with \lmap?] ZMap/LZR performance is
always faster than ZGrab due to LZR's ability to identify service presence
without completing an L7 handshake, which often requires a large number of CPU
cycles for expensive operations (e.g., cryptographic functions in TLS). At
minimum, LZR is 1.9~times faster than ZGrab when scanning 5672/AMQP and, at
maximum, 6.3~times faster when scanning 443/TLS+HTTP\@---equivalent to a 40~CPU
hour speed-up of a 100\% scan of IPv4 when using ZGrab's default number of
senders (1,000) and scanning at ZMap's calculated sending rate that minimizes
ZGrab's packet loss (50K~pps).  The performance of LZR as ZGrab's shim (i.e.,
ZMap/LZR + ZGrab) varies based on a port's service makeup.  When a port contains
a large raw number of hosts that do not consistently establish a TCP connection
(e.g., zero window), there is substantial performance improvement: ZMap/LZR +
ZGrab is 2.6~times faster than ZGrab when scanning 62220/HTTP\@.  On the
contrary, since the \textit{relative} number of hosts that do not consistently
establish a TCP connection on port~443 is small, there is little improvement
(1.1~times).

When a significant fraction of candidate services do not acknowledge data, there
is significant improvement when using \lmap to filter hosts offline (i.e., when
ZGrab can run at full speed). On a 100\% IPv4 scan of 27017/MongoDB, only 21\% of
hosts that \sa acknowledge data and an additional 30\% of hosts send a zero
window, which allows \lmap to increase ZGrab performance by 7~times and
a \lmap scan by 55~times. Unpopular ports are expected to have the
same performance improvement as 62220/HTTP (e.g., a 25~times speed-up) because IPs
on the majority of ports are more likely to not acknowledge data when sending a
\sa.

\paragraph[How much bandwidth does \lmap save?] Using \lmap
alone to fingerprint services always saves bandwidth (up to 87\% on
47808/HTTP+TLS) when the reasonably-expected data is sent during the
initial handshake, as (1) \lmap does not attempt to re-transmit \ack{s} to
zero-window hosts to check for an increase in window size, and  (2) \lmap does
not need to complete full L7 handshakes. However, when using \lmap alongside
ZGrab when scanning a port where the majority of TCP-responsive hosts serve the
expected protocol, there exists an overhead in the number of total packets
sent---even when there is a speed-up in time---due to \lmap sending at least one
extra \ack to fingerprint before re-attempting the actual handshake (e.g., \lmap
+ ZGrab together send 28\% more packets than ZMap+ZGrab for 80/HTTP even though \lmap + ZGrab
run 1.2~times faster than ZMap+ZGrab).

%\subsection{Summary}
%
%\lmap quickly detects and fingerprints both expected and unexpected services
%with often just one extra packet beyond ZMap, thereby greatly increasing the
%efficiency of detecting unexpected services.  LZR can make ZGrab perform up to
%55x faster while discovering 50\% of services that would otherwise require
%running 30~unique Layer~7 handshakes.
%

\section{Related Work}

Fast Internet-wide scanning has been used in hundreds of academic papers in the
past seven years. While we cannot enumerate every paper that has used the
technique, we emphasize that scanning is now common in the security, networking,
and Internet measurement communities. Data collected through Internet-wide scans
has been used to understand censorship~\cite{pearce2017augur, pearce2017global,
khattak2016you}, botnet behavior~\cite{mirai, hunting}, patching
behavior~\cite{li2016you, durumeric2014matter, durumeric2014internet} as well as
to uncover vulnerabilities in IoT and SCADA devices~\cite{costin2014large,
mirian2016internet, springall2016ftp}, cryptographic protocols like
TLS~\cite{amann2017mission, holz2015tls, checkoway2014practical,
aviram2016drown, beurdouche2015messy}, SSH~\cite{adrian2015imperfect,
heninger2012mining}, and SMTP~\cite{durumeric2015neither}, and the Web
PKI~\cite{durumeric2014matter}. Multiple tools have emerged in the space, most
notably ZMap~\cite{durumeric2013zmap} and Masscan~\cite{graham2014masscan}. As
of 2020, more than 300~papers used ZMap and in 2014, Durumeric et~al.\ found
that a significant fraction of all Internet scanning uses
ZMap~\cite{durumeric2014internet}. Prior to the development of these tools in
2013, groups performed smaller-scale studies to measure a multitude of Internet
dynamics (e.g.,~\cite{census}).

Despite the growing popularity of the technique, there has been relatively
little work specifically investigating the dynamics of Internet-wide scanning.
Several works have noted the large discrepancy between L4 and L7
responses~\cite{mirian2016internet, springall2016ftp, heninger2012mining,
durumeric2013zmap, durumeric2013analysis, durumeric2015search}.  Clayton
et~al.~\cite{claytonChinaFirewall} find evidence of dynamic blocking within the
Great Firewall of China---but do not formally quantify how wide-spread the
behavior is---and Wan et~al.~\cite{wan20origins} find evidence of dynamic
blocking within SSH\@.

Alt et~al.\ introduced degreaser~\cite{alt2014uncovering} to locate
``tarpits''---fake services that attempt to trick network scanners; tarpits may
use some of the same techniques we see middleboxes use at the start of a
connection.  In a similar vein to our work, in 2018, Bano
et~al.~\cite{bano2018scanning} studied the notion of host liveness. As part of
their taxonomy, they considered the relationship between live services on
different points, showing that the responses on popular ports are correlated
with one another. In 2014, Durumeric et~al.\ investigated server blacklisting
and how operators respond to Internet-wide scanning; at the time they found that
blacklisting behavior was negligible~\cite{bano2018scanning}. R\"uth et al.
considered the ICMP responses received in response to ZMap IPv4 SYN
scans~\cite{ruth2019hidden}.

One contribution of our work is the introduction of \lmap, which reduces the
time needed to scan less populous ports. Prior work has similarly attempted to
reduce the time required to complete Internet-wide scans, though through starkly
different approaches. Klick et~al.~\cite{klick2016towards} show that much of the
IP address space does not need to be continually scanned by services like
Censys~\cite{durumeric2015search}. Adrian et~al.\ introduce a faster version of
ZMap that operates at 10gbE~\cite{adrian2014zippier}. \lmap solves a different
problem and can be used in coordination with these other performance
improvements. Similar to how we use a single packet to identify services,
several works have focused on single-packet fingerprinting to identify operator
systems~\cite{shamsi2014hershel,shamsi2017faulds}.

\section{Recommendations and Conclusion}
\label{sec:rec_conclusion}

We began our analysis by investigating the troubling observation that a
significant fraction of hosts on the Internet that respond to a SYN scan never
complete an application-layer handshake~\cite{mirian2016internet,
springall2016ftp, heninger2012mining, durumeric2013zmap, durumeric2013analysis,
durumeric2015search}. We found that middleboxes are responsible for the majority
of responses with no real services. We also showed that a significant fraction
of services are also located on unexpected ports. For example, 97\% of HTTP and
93\% of TLS services are not located on ports~80 and 443, respectively.
Worryingly, unexpected services often have weaker security postures than those
on standard ports.

Building on these observations, we introduced \lmap, a scanner that dramatically
reduces the time required to perform an application-layer scan on ports with few
expected services (e.g., 5500\% speedup on 27017/MongoDB) while simultaneously
identifying many unexpected services running on the port. \lmap can identify
16~protocols and 88\% of identifiable services with one packet and 99\% of
identifiable unexpected services with 5~handshakes. Nonetheless, there are two
additional challenges to scanning unassigned ports: (1) scanning 100\% of all
65,535~ports is not feasible, and (2) it is not clear which subset of ports is
worth scanning (e.g., contain a significant fraction of the particular behavior
being studied). We therefore recommend that researchers conduct lightweight
sub-sampled (e.g., 0.1\%) application-layer scans across all ports to detect the
prevalence of targeted protocols. We emphasize that merely using the top $n$
most popular ports is not sufficient to evaluate which ports are most likely to
host particular services, as most protocols are drowned out by the overwhelming
popularity of HTTP and TLS\@. We hope that researchers find \lmap helpful in
accurately and efficiently identifying services in Internet-wide scans.

\section*{Acknowledgements}

The authors thank Tatyana Izhikevich, Katherine Izhikevich, Kimberly Ruth,
Deepak Kumar, David Adrian, Deepti Raghavan, Jeff Cody, members of the Stanford
University and UC San Diego security and networking groups, and the anonymous
reviewers for providing insightful discussion and comments on various versions
of this work.  We further thank Sadjad Fouladi and Katherine Izhikevich for
using their artistic talent to greatly improve the visual graphics in this work.
This work was supported in part by the National Science Foundation under award
CNS-1823192, Cisco Systems, Inc., Google., Inc., the NSF Graduate Fellowship
DGE-1656518 and a Stanford Graduate Fellowship.

%\todo{zakir+renata add thank yous + grants}
{\footnotesize \balance \bibliographystyle{abbrv}
\bibliography{reference}}

\appendix
\onecolumn
\begin{landscape}
\section{Protocols Scanned}
\label{app:protos}
\begin{figure*}[h!]
%\centering
\begin{minipage}[t]{0.5\linewidth}
%\footnotesize
\begin{tabular}{lllllll}
\toprule
	Top 30 & Port & Expected Protocol & IANA-Assigned & Scanner \\ %& SYNACKs & L7 \\
\midrule
x	& 80 & HTTP & HTTP & HTTP \\
x	& 443 & HTTPS & HTTPS & TLS \\
x	& 7547 & CWMP (HTTP) & CWMP (HTTP) & HTTP \\
x	& 22 & SSH & SSH & SSH \\
x	& 30005 & - & - & - \\
x	& 5060 & SIP & SIP & - \\
x	& 21 & FTP & FTP & FTP \\
x	& 25 & SMTP & SMTP & SMTP \\
x	& 2000 & sccp & cisco-sccp & - \\
x	& 8080 & HTTP & HTTP & HTTP \\
x	& 50805 & - & - & - \\
x	& 4567 & HTTP & tram & HTTP \\
x	& 53 & DNS & DNS & DNS (TCP) \\
x	& 49154 & - & - & - \\
x	& 49152 & - & - & - \\
x	& 8081 & - & sunproxyadmin & - \\
x	& 8089 & -& - & - \\
x	& 110 & POP3 & POP3 & POP3 \\
x	& 3306 & MYSQL & MYSQL & MYSQL \\
x	& 8085 & - & - & - \\
x	& 8000 & - & irdmi & - \\
x	& 143 & IMAP & IMAP & IMAP \\
x	& 51005 & - & - & - \\
x	& 3389 & RDP & RDP & RDP \\
x	& 587 & SMTP & submission & SMTP \\
x	& 58000 & - & - & - \\
x	& 993 & IMAPS & IMAPS & IMAPS \\
x	& 995 & POP3S & POP3S & POP3S \\

\bottomrule
 \end{tabular}
 \end{minipage}
 \vspace{0pt}
 \begin{minipage}[t]{0.5\linewidth}
 \begin{tabular}{lllllll}
 \toprule
 	Top 30 & Port & Expected Protocol & IANA-Assigned &  Scanner \\ %& SYNACKs & L7 \\

 \midrule
 x	& 465 & SMTP & SMTP & SMTP \\
 x	& 23 & Telnet & Telnet & Telnet \\
 	& 8443 & HTTPS & pcsync-https & TLS \\
	& 1723 & PPTP & PPTP & PPTP \\
	& 179 & BGP & BGP & - \\
	& 5432 & Postgres & Postgres & Postgres \\
	& 1883 & MQTT & MQTT & MQTT \\
	& 5672 & AMQP & AMQP & AMQP \\
	& 8883 & mqtt & secure-mqtt & mqtt \\
%	& 2525 & telnet & ms-v-worlds & telnet \\
%	& 888 & & accessbuilder & - \\
	& 1521 & Oracle DB & Oracle DB & Oracle DB \\
	& 53194 & - & - & - \\
	& 62220 & - & - & - \\
	& 49227 & - & - & - \\
	& 6379 & redis & redis & redis \\
	& 5900 & VNC & VNC & VNC \\
	& 20000 & DNP3 & DNP3 & DNP3 \\
%	& 161 & SNMP & SNMP & SNMP \\
	& 65535 & - & - & - \\
	& 1433 & mssql & mssql & mssql \\
	& 445 & SMB & SMB & SMB \\
	& 631 & IPP & IPP & IPP \\
	& 6443 & Kubernetes & sun-sr-https & Kubernetes \\
%	& 1911 & & mtp & - \\
	& 623 & IPMI & IPMI & IPMI \\
%	& 162 & & snmptrap & - \\
	& 47808 & - & Bacnet & - \\
	& 27017 & Mongodb & Mongodb  & Mongodb \\
	& 502 & Modbus & Modbus & Modbus \\
	& 102 & Siemens S7 & iso-tsap & Siemens S7 \\
	& 11211 & memcached & memcached & memcached \\
	\\
	 
\bottomrule
\end{tabular}
\end{minipage}
\caption{\textbf{Port Selection}---%
Three categories of ports are scanned: (1) The top 30 ports determined by a \sa scan conducted across all 65K ports of 1\% of IPv4. (2) Ports for which a ZGrab-scanner exists (i.e., to be able to complete the full L7 handshake). (3) A random selection of 5 ephemeral ports. We label the expected service being hosted on the port, as well as the IANA-assigned service.  Note that each of these categories contain overlapping ports.
}
\end{figure*}
\end{landscape}

%80/HTTP, 443/HTTPS, 22/SSH, 7547/HTTP, 22/SSH, 21/FTP, 25/SMTP, 8080/HTTP, 4567/HTTP, 443/TLS, 995/POP3S, 993/IMAPS, 8443/TLS, 465/SMTP, 587/SMTP,53/DNS,110/POP3,143/IMAP, 1723/PPTP, 23/TELNET, 3306/MYSQL, 445/SMB, 5432/POSTGRES, 20000/DNP3,161/SNMP, 5900/VNC, 1833/MQTT, 631/IPP, 6379/redis, 11211/memcached, 27017/Mongodb, 1521/Oracle, 623/IPMI, 5672/AMQP, 102/Siemens, 502/modus, 47808/Bacnet, 6443/Kubernetes, 3389/RDP,  1433/MSSQL, 8081/HTT

%\input{majorRevision_Response}

\end{document}